\documentclass[preprint,aps]{revtex4-2}

\usepackage[utf8]{inputenc} % Ensure your file is saved in UTF-8 encoding
\usepackage{graphicx}
\usepackage{bm}
\usepackage{nicefrac}
\usepackage{array} % Add this line in the preamble if not already included
\usepackage{upgreek}
\usepackage{amsmath}
\usepackage{makecell} % Added to enable the \makecell command
\usepackage{booktabs}
\usepackage{multirow} % Added to support the \multirow command
\usepackage{titlesec}
\usepackage{placeins} 
\usepackage{hyperref}
\usepackage{cleveref}
\usepackage{titlesec}
\usepackage{tabularx}

\titleformat{\section}
{\normalfont\Large\bfseries} 
{\thesection}{1em}{}         

\crefname{figure}{Fig.}{Figs.}
\crefname{equation}{Eq.}{Eqs.}

\begin{document}

\titleformat{\section}
  {\normalfont\large\bfseries}{\thesection}{1em}{}
\titleformat{\subsection}
  {\normalfont\normalsize\bfseries}{\thesubsection}{1em}{}
\titleformat{\subsubsection}
  {\normalfont\normalsize\itshape}{\thesubsubsection}{1em}{}

\renewcommand\thesection{\arabic{section}}
\renewcommand\thesubsection{\thesection.\arabic{subsection}}
\renewcommand\thesubsubsection{\thesubsection.\arabic{subsubsection}}

\title{Expansion of Momentum Space and Full 2\(\pi\) Solid Angle Photoelectron Collection in Laser-Based Angle-Resolved Photoemission Spectroscopy by Applying Sample Bias}

\author{Taimin Miao$^{1,2*}$, Yu Xu$^{1}$, Bo Liang$^{1,2}$,  Wenpei Zhu$^{1,2}$, Neng Cai$^{1,2}$, Mingkai Xu$^{1,2}$, Di Wu$^{1,2}$, Hongze Gu$^{1,2}$, Wenjin Mao$^{1,2}$, Shenjin Zhang$^{3}$,
Fengfeng Zhang$^{3}$, Feng Yang$^{3}$, Zhimin Wang$^{3}$, Qinjun Peng$^{3}$, Zuyan Xu$^{3}$, Zhihai Zhu$^{1}$, Xintong Li$^{1}$, Hanqing Mao$^{1}$, Lin Zhao$^{1}$, Guodong Liu$^{1,2}$ and X. J. Zhou$^{1,2,4*}$}

\affiliation{
\\$^{1}$Beijing National Laboratory for Condensed Matter Physics, Institute of Physics, Chinese Academy of Sciences, Beijing 100190, China
\\$^{2}$University of Chinese Academy of Sciences, Beijing 100049, China
\\$^{3}$Technical Institute of Physics and Chemistry, Chinese Academy of Sciences, Beijing, China
\\$^{4}$Songshan Lake Materials Laboratory, Dongguan 523808, China
\\$^{*}$Corresponding authors: mtm@iphy.ac.cn, xjzhou@iphy.ac.cn
}

\date{\today}

\maketitle

{\bf Angle-resolved photoemission spectroscopy (ARPES) directly probes the energy and momentum of electrons in quantum materials, but conventional setups capture only a small fraction of the full 2\(\pi\) solid angle. This limitation is acute in laser-based ARPES, where the low photon energy restricts momentum space despite ultrahigh resolution. Here we present systematic studies of bias ARPES, where applying a sample bias expands the accessible momentum range and enables full 2\(\pi\) solid angle collection in two dimension using our 6.994,eV laser source. An analytical conversion relation is established and validated to accurately map the detector angle to the emission angle and the electron momentum in two dimensions. A precise approach is developed to determine the sample work function which is critical in the angle-momentum conversion of the bias ARPES experiments.  Energy and angular resolutions are preserved under biases up to 100 V, and minimizing beam size is shown to be crucial. The technique is effective both near normal and off-normal geometries, allowing flexible Brillouin zone access with lower biases. Bias ARPES thus elevates laser ARPES to a new level, extending momentum coverage while retaining high resolution, and is applicable across a broad photon-energy range.}

\section{Introduction}

Angle-resolved photoemission spectroscopy (ARPES) has emerged as a powerful technique in condensed matter physics, providing direct access to the energy and momentum of electrons which are two key physical quantities in characterizing the electronic structure of quantum materials \cite{damascelli_2003_Rev.Mod.Phys.,schrieffer_2007_,sobota_2021_Rev.Mod.Phys.}.  When light is incident on a solid, electrons in the material can absorb the photon energy and be emitted from the surface after overcoming the work function. By detecting the number and energy of photoelectrons emitted along different angles and considering the momentum and energy conversions, the energy and momentum of the electrons inside the material can be measured. Since its inception, the ARPES technique has made significant advancements in energy resolution, momentum resolution, and detection efficiency, and continues to undergo further improvements with the advancement of electron energy analyzers and light sources. In particular, the angle collection of photoelectrons has evolved from the early measurement of one angle at one time (zero-dimensional) to the current simultaneous detection of multiple emission angles along a line (one-dimensional) or across a two-dimensional plane (two-dimensional). The utilization of ultraviolet lasers has greatly elevated the performance of the ARPES technique in achieving super high energy resolution, improved momentum resolution, enhanced bulk sensitivity and high data acquisition efficiency and quality \cite{koralek_2007_Rev.Sci.Instrum.,liu_2008_Rev.Sci.Instrum.,kiss_2008_Rev.Sci.Instrum.,he_2016_Rev.Sci.Instrum.,zhou_2018_Rep.Prog.Phys.}. 

\begin{figure*}[tbp]
  \begin{center}
  \includegraphics[width=0.8\columnwidth,angle=0]{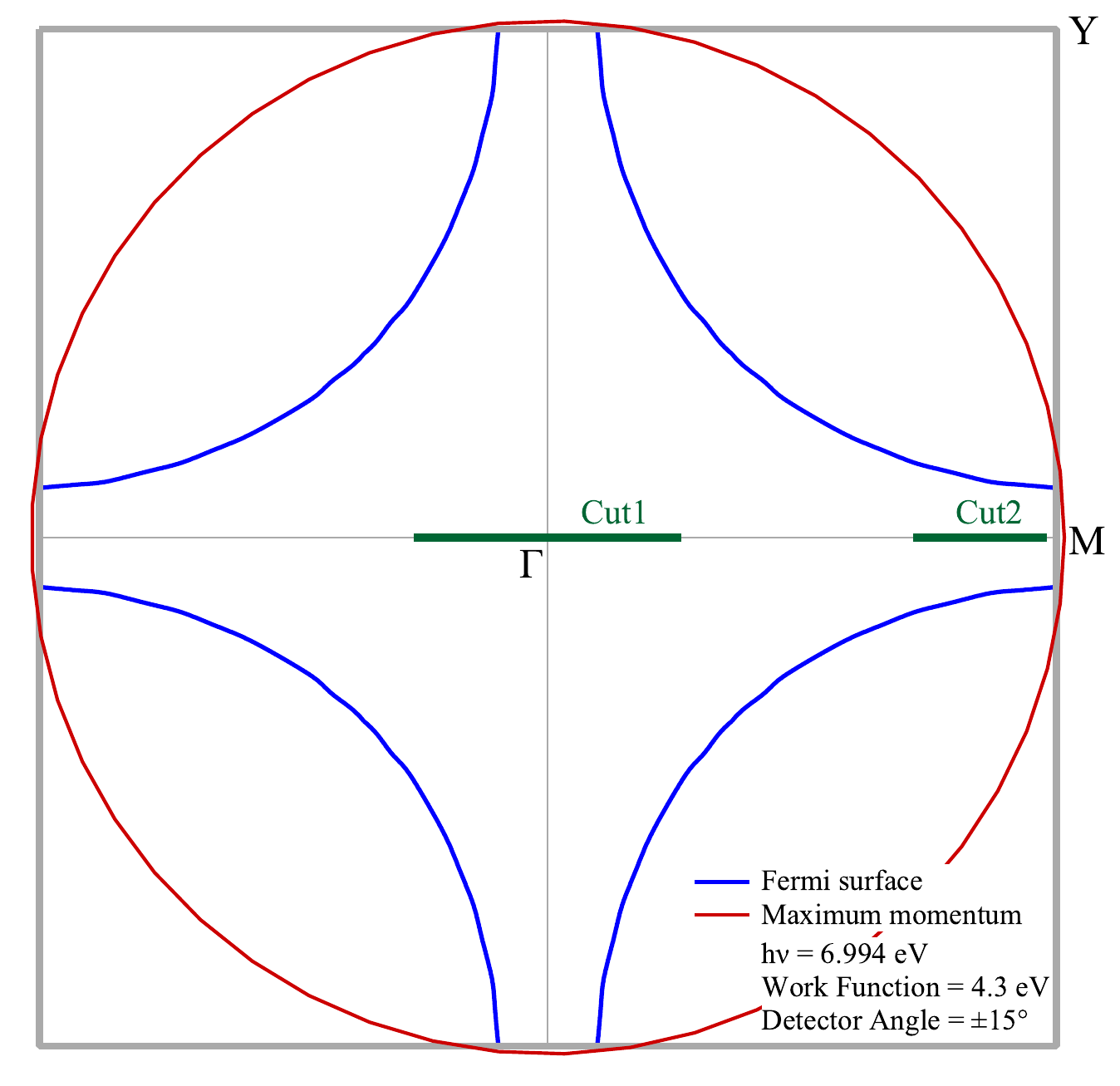}
  \end{center}
  \caption{\textbf{Illustration of the Momentum Coverage in the Laser ARPES Measurements with a Photon Energy $h\nu$=6.994\,eV and the Analyzer Detection Angle of $\pm$15$^\circ$.} Here we take the cuprate superconductors as an example. The gray square represents the first Brillouin zone of the cuprate superconductors. The blue solid lines correspond to typical Fermi surface of cuprate superconductors. The red circle indicates the maximum accessible momentum range when the full 2$\pi$ solid angle of photoelectrons are collected in the  laser ARPES measurements with a photon energy of 6.994\,eV. Here the sample work function of 4.3\,eV is assumed. The green line marked with Cut1 represents the momentum space that can be covered by a single momentum cut with the detector angle of $\pm15^\circ$ near the Brillouin zone center. The other green line marked with Cut2 represents the momentum space that can be covered near the Brillouin zone boundary by the same $\pm15^\circ$ detector angle. It is evident that, to cover the entire Brillouin zone, a lot of momentum cuts need to be carried out and combined together. 
  }
  \label{fig1_motivation}
  \end{figure*}

Despite all these advances, the modern ARPES techniques still experience strong limitations in angular detection of photoelectrons. The mainstream hemispherical electron energy analyzers can collect photoelectrons with an acceptance angle of about $\pm$15$^\circ$ along a line in the real space. Since photoelectrons are emitted across a full 2$\pi$ solid angle around the sample, this represents only a tiny portion of all photoelectrons as the vast majority of photoemitted electrons are not measured. This results in an extremely low detection efficiency. The small acceptance angle of $\pm15^\circ$ gives a limited momentum range along a line in the reciprocal space. As exemplified in \cref{fig1_motivation}, when the vacuum ultraviolet (VUV) laser with a photon energy of 6.994\,eV is used, the momentum range from the $\pm15^\circ$ acceptance angle (Cut1 and Cut2 in \cref{fig1_motivation}) occupies only a small fraction of the entire first Brillouin zone (BZ) of the typical cuprate superconductors. To get access to the entire first BZ, many such momentum line cuts have to be carried out and stitched together. This is time-consuming and it involves multiple mechanical sample rotations, polarization geometry change and sample aging. Consequently, in the ARPES community, it has been a long-standing desire to enhance the collected angular range of photoelectrons and ultimately achieve the full 2$\pi$ solid angle collection.   

It has been found that, by creating an electric field between the sample and analyzer, more photoelectrons are deflected into the analyzer and the emission angle range can be significantly enhanced. In the ARPES mode of the photoemission electron microscopy (PEEM), several kilovolt-per-millimeter electric field are employed which can achieve full 2$\pi$ solid angle collection of photoelectrons\cite{tusche_2015_Ultramicroscopy}. However, the utilization of such a high electric field imposes stringent constraints on the sample surface, thereby limiting the range and types of materials that can be studied. In addition, the limited energy resolution restricts its ability to resolve fine electronic structures. More recently, by applying a moderate bias voltage to the sample, the potential to expand the emission angle range in the conventional ARPES setups has been demonstrated\cite{yamane_2019_Rev.Sci.Instrum.,gauthier_2021_Rev.Sci.Instrum.}. Yamane et al.\cite{yamane_2019_Rev.Sci.Instrum.} showed that, by using a specially designed device to apply a negative bias voltage (-500V) on the sample, the measured emission angle range of the photoelectrons is significantly increased in their synchrotron-based ARPES measurements using a photon energy of 50\,eV. But there were obvious angle distortions in the measured data and the quantitative conversion between the measured angle and the momentum was not established. Later on, Gauthier et al.\cite{gauthier_2021_Rev.Sci.Instrum.} applied a relatively low bias voltage (<60\,V) on the sample in their laser-based ARPES measurements with a photon energy of 6\,eV and observed clear expansion of the collected emission angle range. They successfully established the correspondence between the measured detector angle and the momentum space. Due to the limitation of the electron energy analyzer, their measurements were limited to one-dimensional momentum cuts. It is desirable to extend this approach to expand the two-dimensional momentum space, particularly in the VUV laser-based ARPES systems with super high energy resolution.

In this work, we present a general method of applying a bias voltage on the sample in our VUV laser-based ARPES system (Bias ARPES) with a photon energy of 6.994\,eV, realized by a simple modification of our existing cryostat. By taking advantage of the angle deflection mode in our electron energy analyzer, we successfully achieved full two-dimensional 2\(\pi\) solid angle collection of photoelectrons. To establish a reliable mapping between angle and momentum, we carefully examined the conversion relations between the detector angle, the emission angle and the electron momentum, and demonstrated that the parallel-plate capacitor model provides a reasonable description of the electrical field around the sample, while the position limit properly accounts for the conversion between the detector angle and the emission angle in our laser ARPES system. In addition, we developed a precise approach to determine the sample work function which is critical in the angle-momentum conversion of the bias ARPES experiments.

We investigated the performance of bias ARPES in detail. The application of a bias voltage leads to a minor degradation of the energy resolution, primarily due to the voltage fluctuation from the source meter, but it keeps the high energy resolution (better than 5\,meV) even with the application of a high bias voltage. The angular resolution is also degraded by the sample bias mainly because of the angular magnification and an additional contribution that is dependent on the beam size. We found that the laser beam size is a key factor for achieving sharp spectral features and high instrumental performance in bias ARPES. We further demonstrated that bias ARPES can still work when the sample is tilted off the normal emission direction. This makes it possible to cover one quadrant or part of one quadrant of the Brillouin zone, which significantly reduces the required sample bias voltage and thus improves both the energy and angular resolutions.

\section{Laser ARPES System and Application of Sample Bias}

Our laser-based ARPES system is equipped with a Scienta Omicron DA30L hemispherical electron energy analyzer and utilizes a VUV laser as the light source with a photon energy of 6.994\,eV\cite{liu_2008_Rev.Sci.Instrum.,zhou_2018_Rep.Prog.Phys.}. In the Angular 30 mode, our analyzer can simultaneously detect about $\pm$19.6 degrees along the slit direction. The analyzer can also electrostatically deflect photoelectrons inside the lens perpendicular to the slit by $\pm 15^\circ$ without rotating the sample. The VUV laser is generated via second harmonic generation  by passing the 355\,nm laser through a nonlinear optical crystal, $KBe_2BO_3F_2$. The resulting 177.3\,nm laser is  quasicontinuous-wave with a repetition rate of 80 MHz and a linewidth of 0.26\,meV. In our laser ARPES system, we can continuously vary the laser beam size between 2\,$\mu$m and 1\,mm by using a microlens assembly. The ARPES chamber maintains an ultrahigh vacuum with a base pressure better than 5$\times 10^{-11}$ mbar.

Figure~\ref{fig2_system}a illustrates the schematic layout of the cryostat, sample, VUV laser and the electron energy analyzer in our laser ARPES system. The sample is glued on top of a small sample post which is then screwed into a copper holder. The copper holder is connected with a copper block by a flexible copper braid which facilitates sample rotations. In order to apply a bias voltage on the sample, a thin sapphire piece is inserted between the cold head of the cryostat and the copper block. The sapphire piece serves as an electrical insulator while maintaining efficient thermal conduction. The addition of the sapphire piece causes a slight temperature increase of the sample. Nevertheless, our lowest sample temperature can still reach 11.7\,K using helium cooling. A cable attached to the copper holder is connected to a Keithley 2450 source meter, enabling the application of a well-defined bias voltage. The DA30L electron energy analyzer measures the kinetic energy and angle of photoelectrons that enter the analyzer. When a bias voltage is applied to the sample, an electric field is established between the sample and the analyzer, modifying the trajectories of the emitted photoelectrons. A negative bias bends the photoelectrons toward the lens axis, guiding them into the acceptance cone of the electron energy analyzer. At sufficiently high bias, all photoelectrons within the full \(2\pi\) solid angle can be collected by the analyzer.

\section{Conversion of the Measured Detector Angle into Photoelectron Emission Angle and Electron Momentum}

\begin{figure*}[tbp]
\begin{center}
\includegraphics[width=0.78\columnwidth,angle=0]{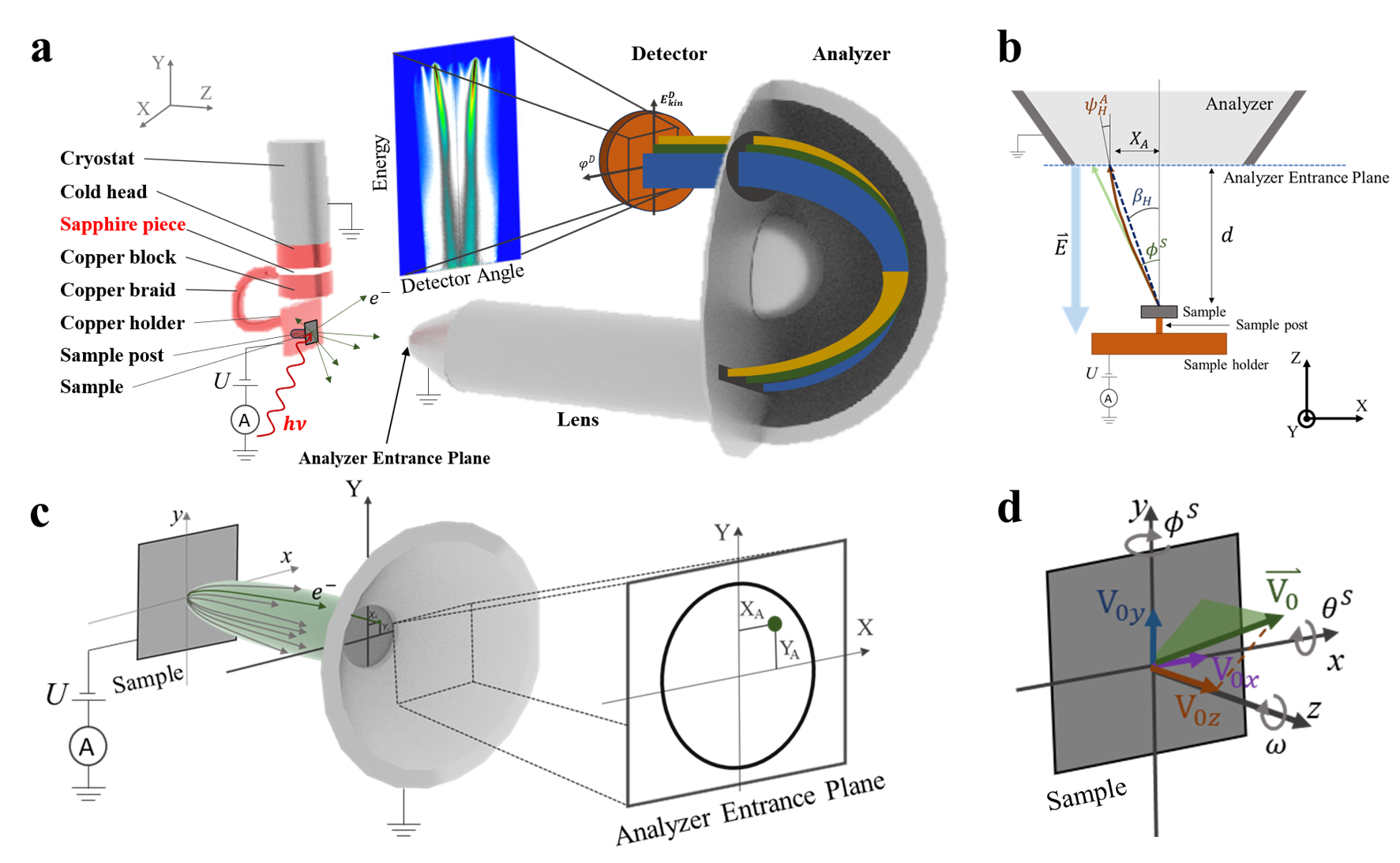}
\end{center}
\caption{\textbf{Schematic Diagram of the Instrument Setup for Bias ARPES Measurement.} {\textbf{a,}} Schematic layout of the cryostat (left side) and the electron energy analyzer (right side). The sample is mounted on top of the sample post which is screwed into a copper holder. The copper holder is connected with another copper block by a flexible copper braid. The copper block is mounted onto the cold head of the cryostat with a sapphire piece in between. The sapphire piece serves to electrically isolate the sample and is thermally conductive. A cable is connected to the sample holder which can either ground the sample or apply bias voltage on the sample. The electron energy analyzer measures the energy and angle of photoelectrons which are emitted from the sample surface along different directions. The detector angle $\varphi^D$ is defined along the angle direction of the detector.
{\textbf{b,}} Schematic view of the bias ARPES in the X-Z plane. The emission angle $\phi^S$, the acceptance angle $\psi^A_H$ and effective emission angle $\beta_H$ are defined. In the Y-Z plane the corresponding angles are $\theta^S$, $\psi^A_V$, and $\beta_V$. {\textbf{c,}} Three-dimensional schematic view of the bias ARPES. When a negative bias voltage is applied to the sample, the photoelectrons are bent towards the lens axis and enter the cone of the electron energy analyzer. When the bias voltage is high enough, all the photoelectrons can be collected into the analyzer. In the analyzer entrance plane, the position of a photoelectron is defined by (X$_A$,Y$_A$) in the X-Y coordinate system. {\textbf{d,}} Definitions of the sample orientation and the velocity of a photoelectron. The sample orientation is determined by three angles $\phi^S$, $\theta^S$ and $\omega$. The velocity of the photoelectron right at the sample surface is represented by {\textbf{\(\vec{V_{0}}\)}} which has three components (V$_{0x}$, V$_{0y}$, V$_{0z}$) in the x-y-z coordinate system.}
\label{fig2_system}
\end{figure*}

ARPES can determine the electron energy ($E$) and momentum ($k$) inside the materials by measuring the kinetic energy and emission angle of the photoemitted electrons. In the photoemission process, the energy and the in-plane momentum are conserved. This leads to the relations:
\begin{align}
  &E_{kin}^S = h\nu - W_S - E_B, \label{eq_PhysicalE}\\
  &k_x = \frac{1}{\hbar} \sqrt{2mE_{kin}^S} \sin\phi^S, \label{eq_PhysicalKx}\\
  &k_y = \frac{1}{\hbar} \sqrt{2mE_{kin}^S} \cos\phi^S \sin\theta^S.\label{eq_PhysicalKy}
\end{align}
where \(E_{kin}^S\) denotes the kinetic energy of the photoemitted electron at the sample surface, \(h\nu\) is the photon energy of the incident light, \(W_S\) represents the sample work function, \(E_B\) is the electron binding energy inside the material, \(\phi^S\) and \(\theta^S\) represent the photoelectron emission angles at the sample surface along two perpendicular x and y directions (\cref{fig2_system}b and \ref{fig2_system}d), \(k_x\) and \(k_y\) represent two electron momentum components inside the material, \(m\) is the electron rest mass and \(\hbar\) is the reduced Planck constant.  In order to determine the energy ($E_B$) and momentum (\(k_x\) and \(k_y\)) inside the material, it is necessary to measure the kinetic energy (\(E_{kin}^S\)) and emission angle (\(\phi^S\) and \(\theta^S\)) of photoemitted electrons at the sample surface. 

As schematically shown in \cref{fig2_system}b and \ref{fig2_system}c, without sample bias, the photoelectron travels to the analyzer along a straight line (green line in \cref{fig2_system}b) with an emission angle of $\phi^S$ ($\theta^S$) in the X-Z (Y-Z) plane and a kinetic energy ($E_{kin}^S$). By applying a negative bias voltage on the sample to form an electric field between the sample and the analyzer, the photoelectron trajectory is deflected towards the analyzer axis along a parabolic trajectory (red line in \cref{fig2_system}b). To clarify the electron trajectory, we define angles, energies and positions at three key reference points. The first is at the sample surface position where the initial emission angles, $\phi^S$ and $\theta^S$, and the effective emission angles, $\beta_H$ and $\beta_V$ (\cref{fig2_system}b), are defined. The initial kinetic energy of photoemitted electrons at the sample surface is denoted as $E_{kin}^S$. The second is on the analyzer entrance plane where the photoelectron position is denoted as $(X_A,Y_A)$ (\cref{fig2_system}b and \ref{fig2_system}c) and the analyzer entrance angles, $\psi_H^A$ and $\psi_V^A$ (\cref{fig2_system}b), are defined. The third is at the analyzer detector position where the detector angle, $\varphi^D$ (along the analyzer slit) and $\vartheta^D$ (perpendicular to the analyzer slit), are defined (\cref{fig2_system}a). The kinetic energy of photoelectrons entering the analyzer is measured and denoted as $E_{kin}^D$. When the sample bias voltage $U$ is applied, given that the detector kinetic energy ($E_{kin}^D$) and the detector angles ($\varphi^D,\vartheta^D$) are measured or known, one central task is to determine the initial kinetic energy ($E_{kin}^S$) and the initial photoelectron emission angles ($\phi^S,\theta^S$) which are necessary to extract the electron energy and  momentum inside the sample. 

In the photoemission process, the kinetic energy of the photoemitted electrons at the sample surface is given by \cref{eq_PhysicalE} which is determined by the photon energy of the incident light ($h\nu$), the sample work function ($W_S$) and the electron energy inside the sample ($E_B$). When the sample bias voltage $U$ is applied and considering the work function of the analyzer ($W_A$), an corrected bias voltage $U^*$ between the sample and the analyzer is defined as:
\begin{equation}
  U^*=U-(W_S-W_A)/e \label{eq_CorBias}
\end{equation}
where \(e\) is the elementary charge. In this case, the kinetic energy of the photoemitted electrons detected by the analyzer ($E_{kin}^D$) can be expressed as:
\begin{align}
    E_{kin}^D&=E_{kin}^S-eU^* \nonumber\\
    &=(h\nu-W_S-E_B)-eU^* \nonumber\\
    &=h\nu-W_A-E_B-eU \label{eq_DetectedE}
\end{align}
It turns out that the kinetic energy of photoemitted electrons detected by the analyzer ($E_{kin}^D$) is independent of the sample work function ($W_S$). From this relation, the electron energy inside the sample ($E_B$) can be determined by the measured kinetic energy of photoelectrons detected by the analyzer ($E_{kin}^D$).

\subsection{Models for Detector Angle-Emission Angle-Momentum Conversions}

It is essential to find the correspondence between the measured detector angles ($\varphi^D$ and $\vartheta^D$) and the initial emission angles of photoelectrons at the sample surface ($\phi^S$ and $\theta^S$). For simplicity, here we discuss the case that the sample surface normal is aligned with the analyzer's lens axis. When the corrected bias voltage $U^*$ is zero, the measured detector angles are equal to the initial emission angles, i.e., $\varphi^D = \phi^S$ and $\vartheta^D = \theta^S$. However, when a sample bias voltage $U$ is applied and $U^*$ is negative, the photoelectron trajectory is deflected towards the analyzer axis, leading to a complex relationship between the measured detector angles and the initial emission angles. Following the model proposed by Gauthier et al.\cite{gauthier_2021_Rev.Sci.Instrum.}, we treat the electrical field in the sample-analyzer region as a parallel-plate capacitor with a separation distance $d$. This framework gives rise to two limiting cases: the position limit where the detector angles are equal to the effective emission angles (i.e., $\varphi^D =\beta_H$,~$\vartheta^D =\beta_V$), and the angular limit where the detector angles equal to the analyzer entrance angles (i.e., $\varphi^D =\psi_H^A$,~$\vartheta^D =\psi_V^A$). In the following, we extend the model to a two-dimensional emission angles which allows us to measure the two-dimensional momentum space by using bias ARPES. 

\subsubsection{Position limit}

In the position limit, the measured detector angles (\(\varphi^D\) and \(\vartheta^D\)) are interpreted as the effective emission angles (\(\beta_H\) and \(\beta_V\)), i.e., $\varphi^D =\beta_H$ and $\vartheta^D =\beta_V$\cite{gauthier_2021_Rev.Sci.Instrum.}. The relationship between the detector angles (\(\varphi^D\) and \( \vartheta^D\)) and the electron entry position (\(X_A \) and \( Y_A\)) (as shown in \cref{fig2_system}b and \ref{fig2_system}c) can then be established as:
\begin{align}
    \varphi^D=\beta_H&=arctan(X_A/d) ,\label{eq_PositionLimitH}\\
    \vartheta^D=\beta_V&=arctan(Y_A/d)\label{eq_PositionLimitV}
\end{align}
The electron entry position, $X_A$ and $Y_A$, is dictated by the traveling process from the sample surface to the analyzer entrance plane. In the process, if we decompose the initial velocity $\vec{V_0}$ of photoelectrons at the sample surface into three components, \( \vec{V_0}=(V_{0x}, V_{0y}, V_{0z})=(V_0\,sin\phi^S, V_0\,cos\phi^S \,sin\theta^S, V_0\,cos\phi^S \,cos\theta^S) \) as indicated in \cref{fig2_system}d, the z component $V_{0z}$ experiences an acceleration due to the electric field ($E=U^*/d$) with an acceleration \( a={-eU^*}/(md) \). The other two components, $V_{0x}$ and $V_{0y}$, keep constant in the process. Considering \cref{eq_PositionLimitH,eq_PositionLimitV}, these give the following relations: 
\begin{align}
    &X_A=V_{0x} \cdot t = V_0\,sin\phi^S \cdot t =d\cdot tan\varphi^D,\label{eq_PositionLimitX}\\
    &Y_A=V_{0y} \cdot t  =V_0\,cos\phi^S \,sin\theta^S  \cdot t =d\cdot tan\vartheta^D,\label{eq_PositionLimitY}\\
    &d=V_{0z} \cdot t+ \frac{1}{2}at^2\label{eq_PositionLimitd}
\end{align}
where \(t\) is the time taken for the photoelectron to travel from the sample surface to the analyzer entrance plane. Note that \cref{eq_PositionLimitX,eq_PositionLimitY} have established the relationship between the measured detector angles ($\varphi^D$ and $\vartheta^D$) and the initial emission angles ($\phi^S$ and $\theta^S$) at the sample surface. However, it involves the time variable $t$ that needs to be solved.

Considering \( V_{0x}^2 + V_{0y}^2 + V_{0z}^2 = V_0^2 \) and substituting $V_{0x},\,V_{0y}$ and $V_{0z}$ from \cref{eq_PositionLimitX,eq_PositionLimitY,eq_PositionLimitd}, one gets the following relation: 
\begin{equation}
 X_A^2+Y_A^2+(d-\frac{1}{2}at^2)^2=V_0^2t^2
\end{equation}
Solving this equation, $t$ can be expressed as:  
\begin{equation}
  t=\frac{\sqrt{2}V_0}{a}\sqrt{\alpha +1\pm\sqrt{2\alpha+1-\alpha^2(tan^2\varphi^D+tan^2\vartheta^D)}}
\end{equation}
where $\alpha$ is a new parameter defined as:
\begin{equation}
  \alpha \equiv \frac{ad}{V_0^2}=\frac{\frac{-eU^*}{md}d}{V_0^2}=-\frac{eU^*}{mV_0^2}=-\frac{eU^*}{2E_{kin}^S} \label{eq_alpha}
\end{equation}

After $t$ is obtained, \cref{eq_PositionLimitX,eq_PositionLimitY} can be rewritten to explicitly describe the relationship between the initial emission angles ($\phi^S$ and $\theta^S$) at the sample surface and the measured detector angles ($\varphi^D$ and $\vartheta^D$): 
\begin{align}
    sin\phi^S &= F_P tan\varphi^D\label{eq_Phi}\\
    cos\phi^S sin\theta^S &= F_P tan\vartheta^D\label{eq_theta}
\end{align}
where the position conversion factor $F_P$ is given by:
\begin{equation}
  F_P=\sqrt{\frac{\alpha +1\mp\sqrt{2\alpha+1-\alpha^2(tan^2\varphi^D+tan^2\vartheta^D)}}{2(1+tan^2\varphi^D+tan^2\vartheta^D)}}\label{eq_Fp}
\end{equation}
In this case, among two solutions, only the  positive branch is taken because the negative branch yields non-physical results in the limiting case of $\alpha \rightarrow 0$, that leads to $F_P \rightarrow 0$.

Finally, the energy ($E_B$) and momentum ($k_x$ and $k_y$) of electrons inside the material can be obtained from the measured kinetic energy ($E_{kin}^D$) and the measured detector angles ($\varphi^D$ and $\vartheta^D$) by considering \cref{eq_DetectedE,eq_Phi,eq_theta} and rewriting \cref{eq_PhysicalE,eq_PhysicalKx,eq_PhysicalKy} as follows: 
\begin{align}
  &E_{kin}^S = E_{kin}^D+eU^*=h\nu - W_S - E_B,\label{eq_Es}\\
  &k_x =\frac{1}{\hbar} \sqrt{2mE_{kin}^S} F_P \tan\varphi^D, \label{eq_kx}\\
  &k_y =\frac{1}{\hbar} \sqrt{2mE_{kin}^S} F_P \tan\vartheta^D\label{eq_ky}
\end{align}

\subsubsection{Angular limit}

In the angular limit, the measured detector angles ($\varphi^D$ and $\vartheta^D$) are interpreted as the analyzer entrance angles (\(\psi_H^A\) and \(\psi_V^A\)), i.e., $\varphi^D =\psi_H^A$ and $\vartheta^D =\psi_V^A$\cite{gauthier_2021_Rev.Sci.Instrum.}. Omitting the detailed derivation as done in the position limit case, we directly present the final results here. The initial emission angles at the sample surface ($\phi^S$ and $\theta^S$) can be expressed in terms of the measured detector angles ($\varphi^D$ and $\vartheta^D$) as follows:
\begin{align}
    &sin\phi^S = F_A \,tan\varphi^D\label{eq_phiA}\\
    &cos\phi^S \,sin\theta^S = F_A \,tan\vartheta^D\label{eq_thetaA}
\end{align}
where the angular conversion factor \(F_A\) is given by:   
\begin{equation}
  F_A=\sqrt{\frac{2\alpha+1}{1+tan^2\varphi^D+tan^2\vartheta^D}}
\end{equation}
in which the parameter \(\alpha\) is the same as expressed in \cref{eq_alpha}. 
The energy ($E_B$) and momentum ($k_x$ and $k_y$) of electrons inside the material can be obtained from the measured kinetic energy ($E_{kin}^D$) and the measured detector angles ($\varphi^D$ and $\vartheta^D$) by considering \cref{eq_DetectedE,eq_phiA,eq_thetaA} and rewriting \cref{eq_PhysicalE,eq_PhysicalKx,eq_PhysicalKy} as follows: 
\begin{align}
    &E_{kin}^S = E_{kin}^D+eU^*=h\nu - W_S - E_B, \\
    &k_x =\frac{1}{\hbar} \sqrt{2mE_{kin}^S} F_A\, \tan\varphi^D, \\
    &k_y =\frac{1}{\hbar} \sqrt{2mE_{kin}^S} F_A \,\tan\vartheta^D
\end{align}

Gauthier et al. found that the position limit is more appropriate to describe their results in their bias ARPES measurements\cite{gauthier_2021_Rev.Sci.Instrum.}. In our case, we also found that the position limit can properly describe our results as presented below.  

\subsection{Spectral Weight Transformation for Detector Angle-Emission Angle-Momentum Conversions}

In ARPES experiments, we measure photoelectron intensity as a function of the detector kinetic energy (\(E_{kin}^D\)) and the detector angles ($\varphi^D$ and $\vartheta^D$), denoted as \( N(E_{kin}^D,\varphi^D,\vartheta^D)\). To describe the electronic structure of materials, these measured results need to be converted to the spectral weight distribution in the emission angle space and further to the momentum space. The spectral weight distribution in the emission angle space is denoted as \( I(E_{kin}^S,\phi^S,\theta^S) \), while the spectral weight distribution in the momentum space as a function of the electron energy ($E_B$) and momentum ($k_x$ and $k_y$) inside the material is denoted as \( A(E_B,k_x,k_y) \). The transformation between the detector angle space, the emission angle space, and the electron momentum space alters the corresponding phase-space volume elements. Specifically, it maps \( \mathrm{d}E_{kin}^D\, \mathrm{d}\varphi^D\, \mathrm{d}\vartheta^D \) to \(\mathrm{d}E_{kin}^S\, \mathrm{d}\phi^S\, \mathrm{d}\theta^S \), then to \(\mathrm{d}E_B\, \mathrm{d}k_x\, \mathrm{d}k_y \), and ultimately establishes the direct relation from \( \mathrm{d}E_{kin}^D\, \mathrm{d}\varphi^D\, \mathrm{d}\vartheta^D \) to \(\mathrm{d}E_B\, \mathrm{d}k_x\, \mathrm{d}k_y \). These transformations, in turn, modify the measured intensity distribution. Consequently, a correction by the Jacobian determinant of the transformation, $J$, is required. 
 
\subsubsection{Spectral Weight Transformation for the Detector Angle-Emission Angle Conversion}

When considering the conversion between the detector angle ($\varphi^D$ and $\vartheta^D$) and the sample emission angle  ($\phi^S$ and $\theta^S$), the intensity correction by the converting Jacobian determinant is given by:
\begin{equation}
    \frac{\partial I(E_{kin}^S,\phi^S,\theta^S)}{\partial E_{kin}^S\partial \phi^S \partial \theta^S}=\frac{\partial N(E_{kin}^D,\varphi^D,\vartheta^D)}{\partial E_{kin}^D \partial \varphi^D \partial \vartheta^D}\cdot J_{D-E}^{-1}\label{eq_J}
\end{equation}
where
\begin{equation}
\renewcommand{\arraystretch}{1.5}
\begin{aligned}
J_{D-E} &=\left|\,\begin{matrix}
\dfrac{\partial E_{kin}^S}{\partial E_{kin}^D } & \dfrac{\partial \phi^S}{\partial E_{kin}^D } & \dfrac{\partial \theta^S}{\partial E_{kin}^D }\\
\dfrac{\partial E_{kin}^S}{\partial \varphi^D } & \dfrac{\partial \phi^S}{\partial \varphi^D } & \dfrac{\partial \theta^S}{\partial \varphi^D }\\
\dfrac{\partial E_{kin}^S}{\partial \vartheta^D } & \dfrac{\partial \phi^S}{\partial \vartheta^D } & \dfrac{\partial \theta^S}{\partial \vartheta^D }
\end{matrix}\,\right|
=\left|\,\begin{matrix}
\dfrac{\partial E_{kin}^S}{\partial E_{kin}^D } & \dfrac{\partial \phi^S}{\partial E_{kin}^D } & \dfrac{\partial \theta^S}{\partial E_{kin}^D }\\
0 & \dfrac{\partial \phi^S}{\partial \varphi^D } & \dfrac{\partial \theta^S}{\partial \varphi^D }\\0 & \dfrac{\partial \phi^S}{\partial \vartheta^D } & \dfrac{\partial \theta^S}{\partial \vartheta^D }
\end{matrix}\,\right|\label{eq_Jacobian}\\
&=\dfrac{\partial E_{kin}^S}{\partial E_{kin}^D } \cdot (\dfrac{\partial \phi^S}{\partial \varphi^D } \cdot \dfrac{\partial \theta^S}{\partial \vartheta^D} - \frac{\partial \theta^S}{\partial \varphi^D } \cdot \frac{\partial \phi^S}{\partial \vartheta^D} )
\end{aligned}
\end{equation}

In the position limit, according to \cref{eq_DetectedE,eq_Phi,eq_theta,eq_Fp}, the individual components of the Jacobian are given by:
\begin{align}
    &\frac{\partial E_{kin}^S}{\partial E_{kin}^D}=1\label{eq_J1}\\
    &\frac{\partial \phi^S}{\partial \varphi^D }=\frac{1}{\sqrt{1-F_P^2\,(tan\varphi^D)^2}}[F_P\cdot sec^2\varphi^D + tan\varphi^D \frac{\partial F_P}{\partial \varphi^D}]\label{eq_J2}\\
    &\frac{\partial \theta^S}{\partial \vartheta^D }=\frac{1}{\sqrt{(cos\varphi^D)^2-F_P^2\,(tan\vartheta^D)^2}}[F_P\cdot sec^2\vartheta^D + tan\vartheta^D \frac{\partial F_P}{\partial \vartheta^D}+F_P tan\vartheta^D\cdot tan\varphi^D\cdot \frac{\partial  \phi^S}{\partial \vartheta^D}]\label{eq_J3}\\
    &\frac{\partial \theta^S}{\partial \varphi^D }=\frac{1}{\sqrt{(cos\varphi^D)^2-F_P^2\,(tan\vartheta^D)^2}}[tan\vartheta^D \frac{\partial F_P}{\partial \varphi^D}+F_P tan\vartheta^D\cdot tan\varphi^D\cdot \frac{\partial  \phi^S}{\partial \varphi^D}]\label{eq_J4}\\
    &\frac{\partial \phi^S}{\partial \vartheta^D }=\frac{1}{\sqrt{1-F_P^2\,(tan\varphi^D)^2}}[tan\varphi^D \frac{\partial F_P}{\partial \vartheta^D}]
 \label{eq_J5}
\end{align}
 where the angular derivatives of \( F_P \) are given by:
\begin{equation}
  \begin{aligned}
    \frac{\partial F_P}{\partial x}&=\frac{-F_P^3\cdot tan x \,sec^2x}{\sqrt{2\alpha+1-\alpha^2 (tan^2\varphi^D+tan^2\vartheta^D)}} \label{eq_J6}
  \end{aligned}
\end{equation}
in which \(x\) represents \(\varphi^D\) or \(\vartheta^D\).

In the angular limit, the correction Jacobian takes a similar form with \(F_P\) replaced by \(F_A\). And its angular derivatives are given by:

\begin{equation*}
    \frac{\partial F_A}{\partial x}=\frac{-F_A\cdot tan\,x\,sec^2\,x}{1+tan^2\varphi^D+tan^2\vartheta^D}
\end{equation*}

\subsubsection{Spectral Weight Transformation for the Emission Angle-Momentum Conversion}

The intensity correction by the Jacobian determinant of the transformation between the emission angle space and the momentum space is given by:
\begin{equation}
    \frac{\partial A(E_B,k_x,k_y)}{\partial E_B \partial k_x \partial k_y}=\frac{\partial I(E_{kin}^S,\phi^S,\theta^S)}{\partial E_{kin}^S \partial \phi^S \partial \theta^S}\cdot J_{E-M}^{-1}
\end{equation}
where
\begin{equation}
\renewcommand{\arraystretch}{1.5}
\begin{aligned}
J_{E-M} &=\left|\,\begin{matrix}
\dfrac{\partial E_B}{\partial E_{kin}^S } & \dfrac{\partial k_x}{\partial E_{kin}^S} & \dfrac{\partial k_y}{\partial E_{kin}^S }\\
\dfrac{\partial E_B}{\partial \phi^S } & \dfrac{\partial k_x}{\partial \phi^S } & \dfrac{\partial k_y}{\partial \phi^S }\\\dfrac{\partial E_B}{\partial \theta^S } & \dfrac{\partial k_x}{\partial \theta^S } & \dfrac{\partial k_y}{\partial \theta^S }
\end{matrix}\,\right|
=\left|\,\begin{matrix}
\dfrac{\partial E_B}{\partial E_{kin}^S} & \dfrac{\partial k_x}{\partial E_{kin}^S } & \dfrac{\partial k_y}{\partial E_{kin}^S }\\
0 & \dfrac{\partial k_x}{\partial \phi^S } & \dfrac{\partial k_y}{\partial \phi^S }\\0 & \dfrac{\partial k_x}{\partial \theta^S } & \dfrac{\partial k_y}{\partial \theta^S }
\end{matrix}\,\right|\label{eq_JacobianE_M}\\
&=\dfrac{\partial E_B}{\partial E_{kin}^S } \cdot (\dfrac{\partial k_x}{\partial \phi^S } \cdot \dfrac{\partial k_y}{\partial \theta^S} - \frac{\partial k_y}{\partial \phi^S } \cdot \frac{\partial k_x}{\partial \theta^S} )
\end{aligned}
\end{equation}

According to \cref{eq_PhysicalE,eq_PhysicalKx,eq_PhysicalKy}, the individual components of the Jacobian are given by:
\begin{align*}
    &\frac{\partial E_B}{\partial E_{kin}^S}=1\\
    &\frac{\partial k_x}{\partial \phi^S }=\frac{\sqrt{2mE_{kin}^S}}{\hbar}[cos(\phi^S)]\\
    &\frac{\partial k_y}{\partial \theta^S }=\frac{\sqrt{2mE_{kin}^S}}{\hbar}[cos(\phi^S)\cdot cos(\theta^S)]\\
    &\frac{\partial k_y}{\partial \phi^S }=\frac{\sqrt{2mE_{kin}^S}}{\hbar}[-sin(\phi^S)\cdot sin(\theta^S)]\\
    &\frac{\partial k_x}{\partial \theta^S }=0
\end{align*}

Substituting the above components into \cref{eq_JacobianE_M}, it gives the Jacobian determinant as:
\begin{equation}
    J_{E-M}=\frac{2mE_{kin}^S}{\hbar^2}cos^2(\phi^S)cos(\theta^S)\label{eq_JacobianMS}
\end{equation}

\subsubsection{Spectral Weight Transformation for the Detector Angle-Momentum Conversion}

The mapping from the detector angle (\(\varphi^D\) and \(\vartheta^D\)) to the momentum (\(k_x\) and \(k_y\)) requires another Jacobian correction:
\begin{equation}
    \frac{\partial I(E_B,k_x,k_y)}{\partial E_B \partial k_x \partial k_y}=\frac{\partial N(E_{kin}^D,\varphi^D,\vartheta^D)}{\partial E_{kin}^D \partial \varphi^D \partial \vartheta^D}\cdot J_{D-M}^{-1}
\end{equation}
where,
\begin{equation}
\renewcommand{\arraystretch}{1.5}
\begin{aligned}
J_{D-M} &=\left|\,\begin{matrix}
\dfrac{\partial E_B}{\partial E_{kin}^D } & \dfrac{\partial k_x}{\partial E_{kin}^D } & \dfrac{\partial k_y}{\partial E_{kin}^D }\\
\dfrac{\partial E_B}{\partial \varphi^D } & \dfrac{\partial k_x}{\partial \varphi^D } & \dfrac{\partial k_y}{\partial \varphi^D }\\
\dfrac{\partial E_B}{\partial \vartheta^D } & \dfrac{\partial k_x}{\partial \vartheta^D } & \dfrac{\partial k_y}{\partial \vartheta^D }
\end{matrix}\,\right|
=\left|\,\begin{matrix}
\dfrac{\partial E_B}{\partial E_{kin}^D } & \dfrac{\partial k_x}{\partial E_{kin}^D } & \dfrac{\partial k_y}{\partial E_{kin}^D }\\
0 & \dfrac{\partial k_x}{\partial \varphi^D } & \dfrac{\partial k_y}{\partial \varphi^D }\\0 & \dfrac{\partial k_x}{\partial \vartheta^D } & \dfrac{\partial k_y}{\partial \vartheta^D }
\end{matrix}\,\right|\\
&=\dfrac{\partial E_B}{\partial E_{kin}^D } \cdot (\dfrac{\partial k_x}{\partial \varphi^D } \cdot \dfrac{\partial k_y}{\partial \vartheta^D} - \frac{\partial k_y}{\partial \varphi^D } \cdot \frac{\partial k_x}{\partial \vartheta^D} )\\
\end{aligned}
\end{equation}

In the position limit, according to \cref{eq_Es,eq_kx,eq_ky}, the individual components of the Jacobian are given by:
\begin{align}
    &\frac{\partial E_B}{\partial E_{kin}^D}=1\\
    &\frac{\partial k_x}{\partial \varphi^D }=\frac{\sqrt{2mE_{kin}^S}}{\hbar}[F_P\cdot sec^2\varphi^D + tan\varphi^D \frac{\partial F_P}{\partial \varphi^D}]\\
    &\frac{\partial k_y}{\partial \vartheta^D }=\frac{\sqrt{2mE_{kin}^S}}{\hbar}[F_P\cdot sec^2\vartheta^D + tan\vartheta^D \frac{\partial F_P}{\partial \vartheta^D}]\\
    &\frac{\partial k_y}{\partial \varphi^D }=\frac{\sqrt{2mE_{kin}^S}}{\hbar}\cdot tan\vartheta^D \frac{\partial F_P}{\partial \varphi^D}\\
    &\frac{\partial k_x}{\partial \vartheta^D }=\frac{\sqrt{2mE_{kin}^S}}{\hbar}\cdot tan\varphi^D \frac{\partial F_P}{\partial \vartheta^D}
\end{align}
 where the angular derivatives of \( F_P \) are given by:
\begin{equation}
  \begin{aligned}
    \frac{\partial F_P}{\partial x}&=\frac{-F_P^3\cdot tan x \,sec^2x}{\sqrt{2\alpha+1-\alpha^2 (tan^2\varphi^D+tan^2\vartheta^D)}}
  \end{aligned}
\end{equation}
in which \(x\) represents \(\varphi^D\) or \(\vartheta^D\).

In the angular limit, the same expressions hold with \(F_P\) being  replaced by \(F_A\).

\subsection{Simulated Relationship between the Collected Emission Angle and the Sample Bias}

\begin{figure*}[tbp]
    \begin{center}
    \includegraphics[width=1.0\columnwidth,angle=0]{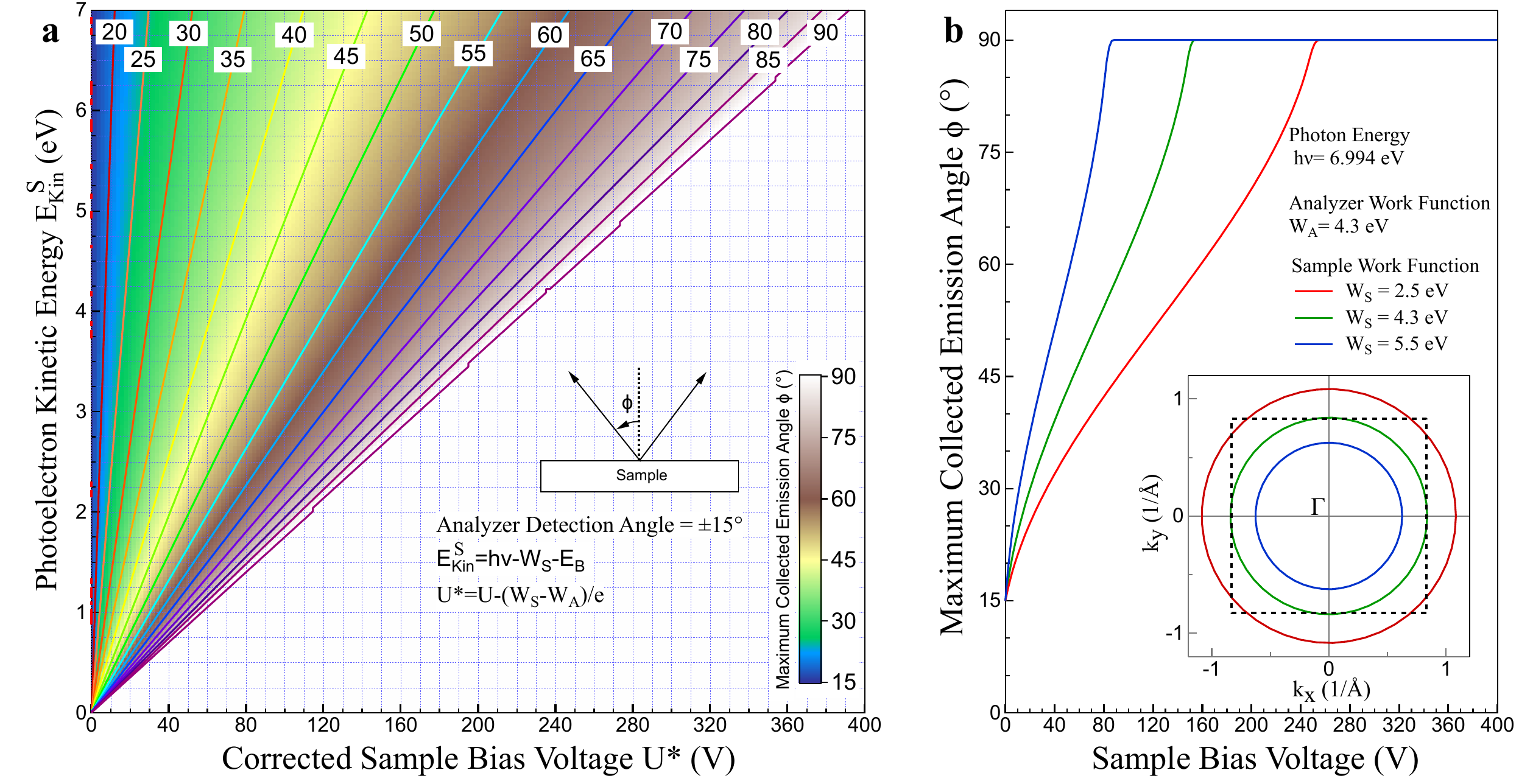}
    \end{center}
    \caption{\textbf{The Maximum Emission Angle of Photoemitted Electrons that can be Collected at Different Bias Voltages.} {\textbf{a,}} Calculated color plot showing the maximum emission angle that can be collected by applying different bias voltages (horizontal axis) on the sample with different work functions (vertical axis). The plot is calculated based on the condition that the photon energy is 6.994\,eV, the analyzer work function is 4.3\,eV and the analyzer detection angle is $\pm$15$^\circ$. The maximum collected emission angle is represented by both the color and lines of constant emission angle. {\textbf{b,}} The  maximum collected emission angle as a function of the sample bias voltage for three typical sample work functions of 2.5\,eV (red line), 4.3\,eV (green line) and 5.5\,eV (blue line). The inset shows the maximum momentum coverage for the three work functions when the entire 2$\pi$ solid angle of photoelectrons is collected. The dashed square represents the first Brillouin zone of cuprate superconductors with a lattice constant $a=3.8$\,\text{\AA}.
    }
    \label{fig3_simulation}
\end{figure*}

We simulated the relationship between the collected emission angle of photoelectrons and the applied sample bias voltage based on the conversion relations \cref{eq_Phi,eq_theta,eq_Fp} for the position limit. Here, the photon energy is taken at $h\nu=6.994\,eV$, the analyzer work function is taken at $W_A=4.3621\,eV$ \cite{huang_2020_Rev.Sci.Instrum.} and the analyzer detection angle range is assumed as $\pm 15^\circ$. The simulated results are presented in \cref{fig3_simulation}a. The horizontal axis denotes the corrected bias voltage \(U^*\) applied to the sample (ranging from 0 to 400 V), while the vertical axis corresponds to the photoelectron kinetic energy \(E_{kin}^S\) (spanning from 0 to 7 eV). For convenience, here and in the following, we display the negative sample bias voltage \(U\) and the negative corrected bias voltage \(U^*\) in their absolute values (\(|U|\) and \(|U^*|\)). The false-color map in \cref{fig3_simulation}a encodes the maximum collectable emission angle (up to 90$^\circ$) as a function of the kinetic energy and the corrected bias voltage. Contour lines of constant maximum emission angle are superimposed on the color map, illustrating how the photoelectron kinetic energy and corrected bias voltage determine the attainable angular coverage. For example, when the photoelectron kinetic energy is 2.7\,eV, a bias voltage of approximately 70\,V is required to achieve a collected emission angle of $\pm$50$^\circ$ and a bias voltage of \(\sim\)155\,V is required to collect the entire $\pm90^\circ$ emission angle. 

Figure~\ref{fig3_simulation}b displays the dependence of the maximum collected emission angle on the sample bias voltage for three typical sample work functions of 2.5\,eV (red curve), 4.3\,eV (green curve) and 5.5\,eV (blue curve). These curves clearly demonstrate the nonlinear dependence of the collected emission angle on the applied voltage, particularly at low bias. This nonlinearity originates from the complex conversion between the detector and emission angles. Notably, samples with higher work functions require smaller bias voltages to achieve full $\pm90^\circ$ collection. For instance, a work function of 2.5 eV corresponds to a required bias of $\sim$255 V, whereas a work function of 5.5 eV reduces this value to $\sim$87 V. However, higher work functions also lead to reduced maximum momentum coverage, as shown in the inset of \cref{fig3_simulation}b. 

\begin{figure*}[tbp]
    \begin{center}
    \includegraphics[width=1.0\columnwidth,angle=0]{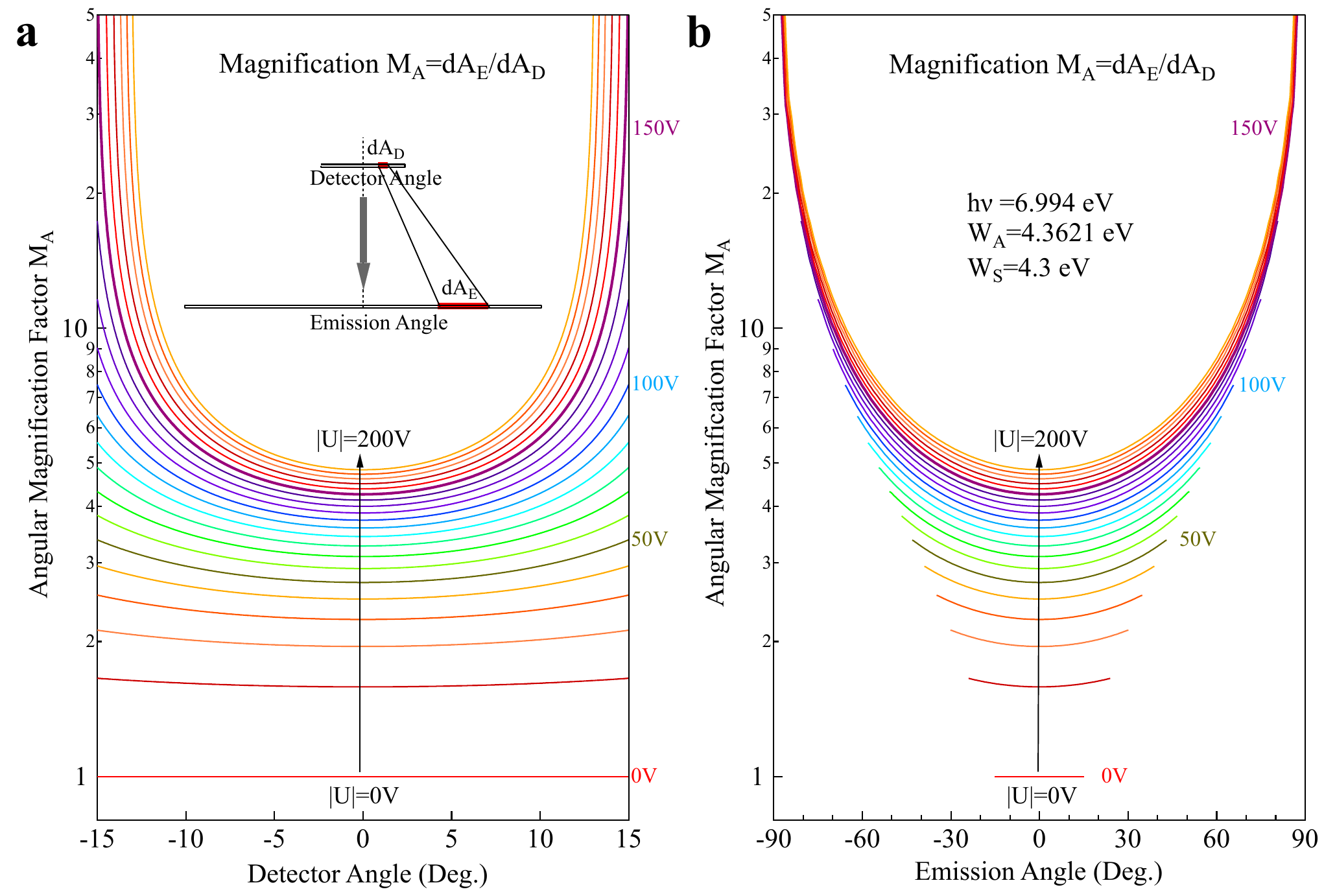}
    \end{center}
    \caption{\textbf{The Angular Magnification Factor (\(M_A\)) between the Detector Angle and the Emission Angle under different Bias Voltages.} A small region (\(dA_D\)) at the  detector angle \(A_D\) is mapped into a region (\(dA_E\)) at the emission angle \(A_E\) as shown in the inset in {\textbf{a}}. The angular magnification factor (\(M_A\)) is defined as \(d A_E /d A_D\) for the detector angle \(A_D\) and the emission angle \(A_E\). {\textbf{a,}} Calculated angular magnification factor (\(M_A\)) as a function of the detector angle under different bias voltages which is varied every 10 Volts. {\textbf{b,}} Calculated angular magnification factor (\(M_A\)) as a function of the emission angle. Here, the photon energy is $h\nu=6.994\,eV$, the analyzer work function is $W_A=4.3621\,eV$ and the sample work function is taken as $W_S=4.3\,eV$.  %These calculations demonstrate that 
    }
    \label{fig4_magnification}
\end{figure*}

In the bias ARPES measurements, in addition to the conversions between the detector angle, the emission angle and the momentum, the spectral weight transformation associated with the angle-momentum conversions is also very important. Such a spectral weight transformation can be realized by \cref{eq_J,eq_Jacobian,eq_J1,eq_J2,eq_J3,eq_J4,eq_J5,eq_J6} in the position limit. The spectral weight transformation between the detector angle and the emission angle is closely related to the angular magnification factor (\(M_A\)). As shown in the inset of \cref{fig4_magnification}a, when a small region (\(dA_D\)) at the  detector angle \(A_D\) is mapped into a region (\(dA_E\)) at the emission angle \(A_E\), the angular magnification factor (\(M_A\)) can be defined as \(M_A=dA_E/dA_D\). This angular magnification factor is crucial for understanding how the angular resolution and intensity distribution are modified during the conversion process.

The angular magnification factor (\(M_A\)) between the detector angle and the emission angle is calculated for different bias voltages and presented in \cref{fig4_magnification}a as a function of the detector angle and in \cref{fig4_magnification}b as a function of the emission angle. Here, the photon energy is $h\nu=6.994\,eV$, the analyzer work function is $W_A=4.3621\,eV$ and the sample work function is taken as $W_S=4.3\,eV$. Under these conditions, a sample bias voltage of 155\,V is required to collect the entire 2\(\pi\) solid angle of photoelectrons. Without applying the sample bias (U=0\,V), the mapped region (\(dA_E\)) in emission angle equals to the original region (\(dA_D\)) in detector angle, and the angular magnification factor (\(M_A\)) is 1 across the entire angular range.  As the applied sample bias voltage increases, the overall magnification increases across the entire angular range. For each bias voltage, the magnification remains nearly constant around the normal emission region but rises at larger angles. This effect gets more pronounced at high bias voltages when it approaches the critical bias to collect the full 2\(\pi\) solid angle (155\,V). For instance, at |U|=150\,V, the magnification factor is 4.3 at the normal emission, slightly increases to 5.7 up to the detector angle of \(\pm 10^\circ\) and then sharply rises to larger than 45 at the edge in the detector angle. This progressive enhancement demonstrates that higher bias voltages not only extend the accessible angular range but also amplify the nonlinear angular expansion toward the edge. The strong angular dependence of the magnification factor at high bias voltage will have a significant effect on the effective angular resolution of the measurements. Consequently, careful consideration of this effect is essential for reliable and quantitative ARPES analysis.

\begin{figure*}[tbp]
    \begin{center}
    \includegraphics[width=1.0\columnwidth,angle=0]{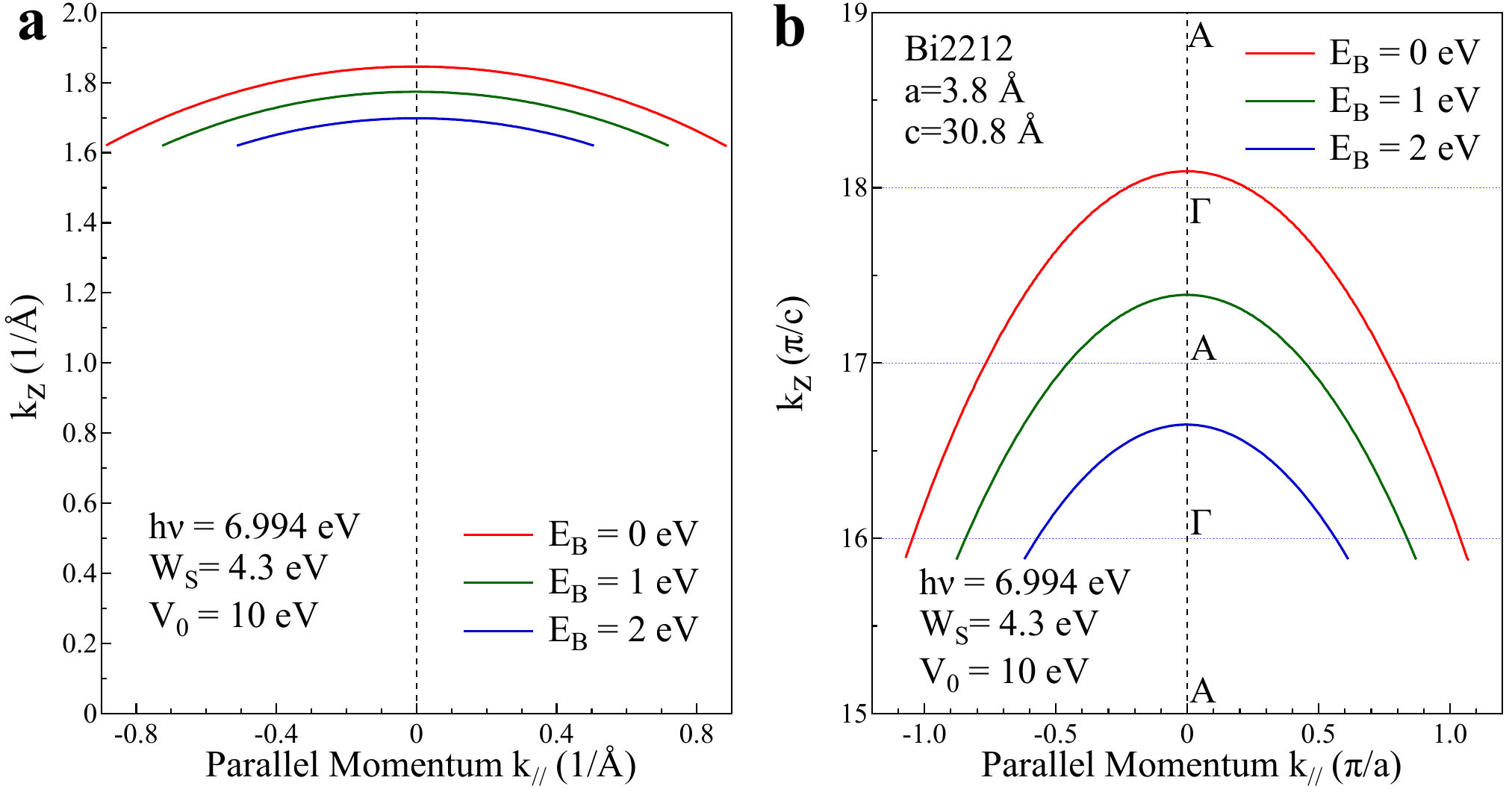}
    \end{center}
    \caption{\textbf{The Calculated $k_z$ Values as a Function of the Parallel Momentum for Different Binding Energies.} \textbf{a,}  Calculated $k_z$ values as a function of the parallel momentum in the absolute momentum unit of 1/\AA for different binding energies of 0 (red), 1 (green) and 2\,eV (blue). Here the photon energy $h\nu$ is 6.994\,eV, the sample work function $W_S$ is 4.3\,eV and the inner potential V$_0$ is taken as 10\,eV. \textbf{b,} Calculated $k_z$ values as a function of the parallel momentum for $Bi_2Sr_2CaCu_2O_8$ (Bi2212) in the relative momentum unit. Here the photon energy $h\nu$ is 6.994\,eV, the sample work function $W_S$ is 4.3\,eV, the inner potential V$_0$ is taken as 10\,eV and the lattice constant along the c-axis is taken as 30.8\,\text{\AA} and a-axis is taken as 3.8\,\text{\AA}. 
    }
    \label{figA_calculatedKz}
\end{figure*}

It is important to recognize that the out-of-plane momentum, $k_z$, may vary with the parallel momentum and the binding energy in the ARPES experiments\cite{damascelli_2003_Rev.Mod.Phys.,borisenko_2025_}. Such an effect becomes more pronounced when the photon energy is low and the covered momentum space becomes large with the application of sample bias. The out-of-plane momentum $k_z$ of photoelectrons inside the material can be estimated by considering the energy and momentum conservations during the photoemission process\cite{damascelli_2003_Rev.Mod.Phys.}:
\begin{equation}    
    k_z =\frac{1}{\hbar} \sqrt{2m(E_{kin}^S cos\phi^S \,cos\theta^S + V_0)} \label{eq_kz_final}
\end{equation}
where \(V_0\) is the inner potential of the material, typically ranging from 10 to 15\,eV for most materials. 
To check the $k_z$ variation in our laser ARPES measurements with the collection of photoelectrons in the full 2$\pi$ solid angle, we performed calculations based on \cref{eq_kz_final} and the results are presented in Fig.~\ref{figA_calculatedKz}a. The results reveal that the $k_z$ variation is significant and cannot be neglected, particularly at low photon energies. Coming to the particular material like Bi\(_2\)Sr\(_2\)CaCu\(_2\)O\(_8\) (Bi2212) (Fig.~\ref{figA_calculatedKz}b), the $k_z$ variation can span the entire Brillouin zone along the $c$-axis.The significant $k_z$ variation inherent to laser ARPES must be carefully considered in data analysis, particularly for materials with pronounced three-dimensional electronic structures.

\section{Determination of Sample Work Function}

\subsection{ARPES Measurement of Sample Work Function} 

The work function is a fundamental quantity of the material surface property which is usually defined as the minium work required to remove electrons from the material to vacuum\cite{halas_2006_Mater.Sci.-Pol.,lin_2023_Phys.Rev.Appl.}. The work function also plays a critical role in the photoemission process, as the measurements of the electron energy and momentum are closely related to the work functions of the analyzer ($W_D$) and/or the sample ($W_S$). The analyzer work function is determined by its components which is basically a constant. It can be determined directly from the difference between the photon energy ($h\nu$) and the measured kinetic energy of photoelectrons at the Fermi level (shown as \cref{eq_DetectedE}). The work function  of our analyzer is also determined precisely by using different photon energies which gives $W_A=4.3621\,eV$\cite{huang_2020_Rev.Sci.Instrum.}. The sample work function can vary widely across materials, typically ranging from \(\sim\)1 to 6 eV\cite{halas_2006_Mater.Sci.-Pol.,lin_2023_Phys.Rev.Appl.}. It plays a crucial role in determining the electron momentum in ARPES measurements, as expressed in \cref{eq_PhysicalE,eq_PhysicalKx,eq_PhysicalKy}, and is particularly significant for laser ARPES using relatively low photon energies. As shown in \cref{eq_CorBias}, even without sample bias (\(U = 0\)), an internal electric field can still arise due to the work function difference between the sample and the analyzer, thereby altering the trajectory of photoelectrons. An accurate determination of the sample work function is required to precisely obtain the electron momentum inside the material. This is particularly significant for laser-based ARPES, where the photon energy is typically low and the work function can dramatically affect the angle-momentum conversion.

Photoemission spectroscopy provides a powerful method for determining the work function of materials. Photoelectrons emitted from the sample surface span an initial kinetic energy range from 0 to $h\nu - W_S$. Considering the work function difference between the sample and the analyzer, the energy window of electrons entering the analyzer is shifted by $(W_S - W_A)$, from $[0,\,h\nu - W_S]$ at the sample surface to $[W_S - W_A,\,h\nu - W_A]$ at the analyzer. If the sample work function $W_S$ is smaller than that of the analyzer $W_A$, part of the low-energy electrons cannot enter the analyzer and are not detected. To overcome this limitation, a negative sample bias $U$ is applied, which shifts the measured kinetic-energy range to $[W_S - W_A + e|U|,\,h\nu - W_A + e|U|]$. In the measured photoemission spectrum (energy distribution curves, EDCs), the Fermi level $E_F$ appears at the high-energy edge $(h\nu - W_A + e|U|)$, while the secondary-electron cutoff $E_{SEC}$ appears at the low-energy edge $(W_S - W_A + e|U|)$. Their separation, $E_F - E_{SEC} = h\nu - W_S$, allows the sample work function to be determined as $W_S = h\nu - (E_F - E_{SEC})$. Therefore, to precisely determine the sample work function, it is necessary to apply a negative sample bias, particularly for materials with low work function. It is also crucial to accurately measure the secondary electron cutoff and the Fermi level in the photoemission spectrum.

\begin{figure*}[tbp]
\begin{center}
\includegraphics[width=1.0\columnwidth,angle=0]{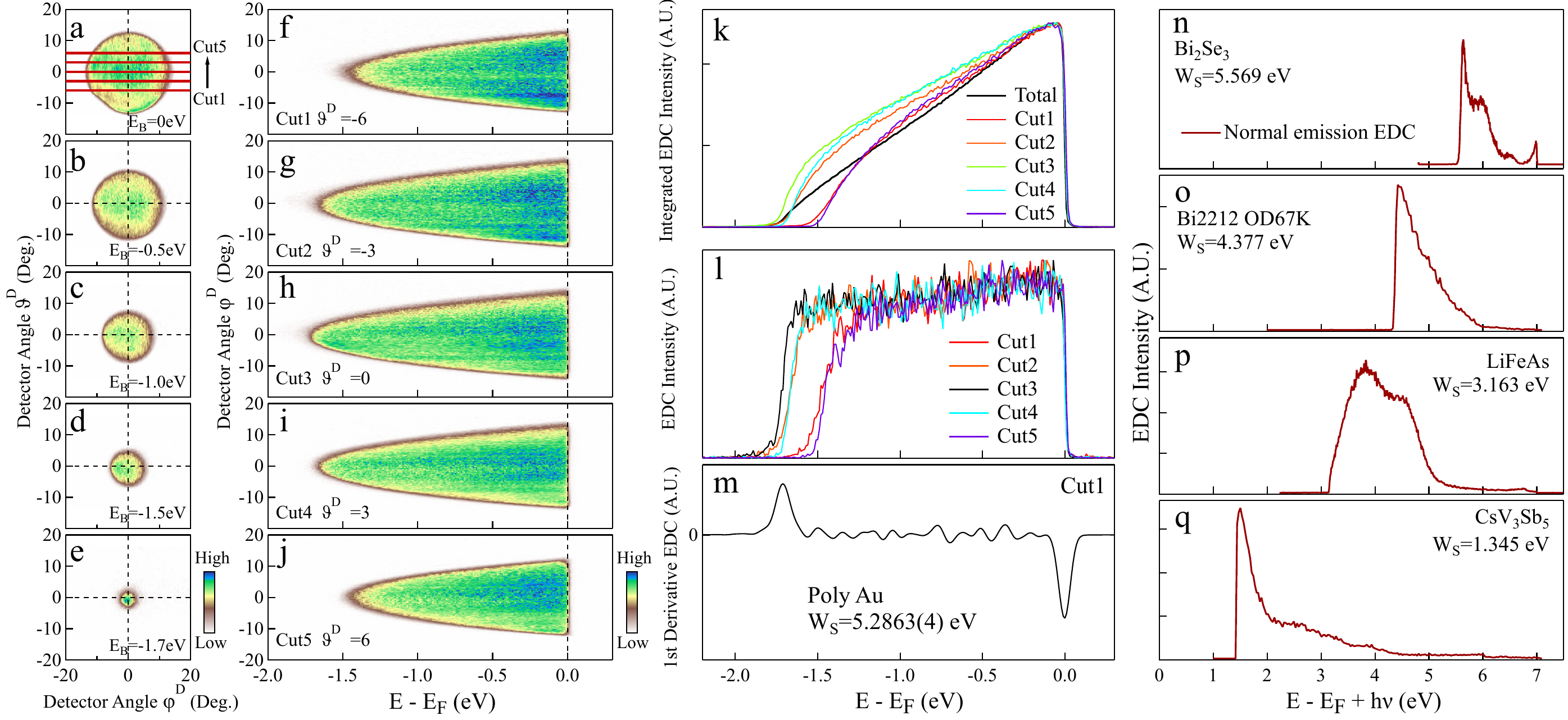}
\end{center}
\caption{\textbf{Sample work function measurements by using the normal emission photoemission spectrum. The photon energy h$\nu$ is 6.994\,eV.} {\textbf{a-e,}} Constant energy contours of polycrystalline gold measured at a bias voltage |U|=90\,V, at the binding energy (E$_B$) of 0\,eV (a), 0.5\,eV (b), 1.0\,eV (c) , 1.5\,eV (d) and 1.7\,eV (e). The spectral intensity progressively concentrates to the normal emission region with increasing binding energy. {\textbf{f-j,}} Photoemission images taken along five cuts at different angle $\vartheta^D$. The location of these cuts is shown by red lines in (a).  The photoelectrons form a cone in the energy-angle space, with the cone bottom at the normal emission. {\textbf{k,}} Integrated photoemission spectra (energy distribution curves, EDCs) obtained from (f-j) by integrating the spectral intensity over all the detector angle $\varphi^D$. The total intensity is also presented (black line) which is obtained by integrating all photoelectrons over the detector angles $\varphi^D$ and $\vartheta^D$. In this case, the high binding energy cutoffs of the secondary electrons are broad and not well-defined.  {\textbf{l,}} EDCs obtained from (f-j) at the detector angle $\varphi^D=0$. In this case, the high binding energy cutoffs of the secondary electrons are sharp and well-defined.  {\textbf{m,}} The first derivative of the EDC at the normal emission (the black curve in (l) for Cut~3). The left peak corresponds to the secondary electron cutoff while the right peak represents the Fermi level cutoff. The precise position of these two peaks are used to determine the sample work function. {\textbf{n-q,}} Normal emission EDCs measured on some representative samples, including Bi\(_2\)Se\(_3\)(n), overdoped Bi2212 with a Tc of 67\,K (o), LiFeAs(p) and CsV\(_3\)Sb\(_5\) (q). The horizontal axis is plotted as E-E$_F$+h$\nu$ so the sample work function can be directly obtained from the position of the secondary electron cutoff.
}
\label{fig5_workfunction}
\end{figure*}

Figures~\ref{fig5_workfunction}a-e present the constant energy contours of polycrystalline gold at different binding energies. During the measurements, a negative sample bias voltage of 90\,V was applied. Figs.~\ref{fig5_workfunction}f-j display photoemission images along five angle cuts with the cut positions marked in \cref{fig5_workfunction}a. With the bias applied, the photoelectrons form a cone in the energy-detector angle ($E$, $\varphi^D$, $\vartheta^D$) space. The cone bottom corresponds to the cutoff of the secondary electron along the normal emission. 

Figure~\ref{fig5_workfunction}k shows the angle integrated EDCs from Figs.~\ref{fig5_workfunction}f-j and the total EDC from integrating over the entire electron cone (black line in \cref{fig5_workfunction}k). While there are clear Fermi level cutoffs on the high-energy side, the secondary electron cutoffs on the low-energy side appear significantly broadened. This makes it difficult to accurately determine its energy position. In contrast, \cref{fig5_workfunction}l shows EDCs extracted from the zero detector angle $\varphi^D$ from \cref{fig5_workfunction}f-j which display sharp and well-defined secondary electron cutoffs. We emphasize here that the sample work function should be determined from the EDC along the normal emission. Therefore, it is crucial to locate precisely the position of the normal emission. This is accomplished by the constant energy contours at high binding energy which will shrink into a point at the cutoff energy (\cref{fig5_workfunction}e). Among the five cuts, Cut~3 corresponds to this normal emission, as shown by the black curve in Fig.~\ref{fig5_workfunction}l. To locate the two cutoff energies unambiguously, we take the first derivative of the EDC from Cut~3 [Fig.~\ref{fig5_workfunction}m]. Two distinct peaks emerge: the left peak marks the secondary-electron cutoff, while the right corresponds to the Fermi level. Their energy separation directly yields the sample work function, giving $W_S = 5.2863(4)$,eV for polycrystalline gold.

By using this method, we have measured the work function of many materials. Figs.~\ref{fig5_workfunction}n-q highlight the measured EDCs along normal emission for several prototypical materials, including Bi\(_2\)Se\(_3\) (\cref{fig5_workfunction}n), overdoped Bi2212 (\cref{fig5_workfunction}o), LiFeAs (\cref{fig5_workfunction}p) and CsV\(_3\)Sb\(_5\) (\cref{fig5_workfunction}q). We have measured the work function of many materials and the results are listed in Table~I.

\subsection{Understanding the Photoelectron Cone Boundary in the Energy-Detector Angle Space} 

As shown in \cref{fig5_workfunction}h, the photoelectron distribution in the energy-detector angle ($E$, $\varphi^D$, $\vartheta^D$) space forms a cone, whose bottom corresponds to the secondary-electron cutoff at normal emission. The cone boundary arises from the angular distribution of photoelectrons emitted from the sample surface: electrons at the cone bottom have zero kinetic energy ($E_{kin}^S$) at the surface, whereas those at the Fermi level possess the highest kinetic energy.  

According to \cref{fig3_simulation}, the corrected sample bias voltage ($U^*$) required to collect the full $2\pi$ solid angle strongly depends on the photoelectron kinetic energy. A relatively small bias (e.g., $\sim 12$\,V for 0.2\,eV electrons) suffices at low kinetic energies, while a much larger bias (e.g., $\sim 155$\,V for 2.7\,eV electrons) is needed at higher energies. The minimum bias necessary to encompass the $2\pi$ solid angle at the Fermi level defines a critical value, $U^*_C$, above which all photoemitted electrons are collected. Under such conditions, the photoelectron cone boundary directly maps the contour of $2\pi$ angular collection at different binding energies.

The photoelectrons with the emission angle of 90\(^\circ\) have zero velocity component along the surface normal (the z-direction), i.e., \(V_{0z}=0\). The initial velocity of these photoelectrons becomes \( \vec{V_0}=(V_{0x}, V_{0y}, V_{0z})=(V_0\,sin\phi^S, V_0\,cos\phi^S \,sin\theta^S, V_0\,cos\phi^S \,cos\theta^S)= (V_0\,sin\phi^S, V_0\,cos\phi^S \,sin\theta^S, 0)\).  This leads to the emission angle relation:
\begin{equation}
  (sin\phi^S)^2+(cos\phi^S \,sin\theta^S)^2=1\label{eq_Vz0}
\end{equation}

By considering the relation between the emission angle and the detector angle in the position limit, \cref{eq_Phi,eq_theta}, the two detector angles are related by: 
\begin{equation}
    (F_P \,tan\varphi^D)^2+(F_P \,tan\vartheta^D)^2=1
\end{equation}
Furthermore, by applying \cref{eq_alpha,eq_Fp}, one gets a simple detector angle relation: 
\begin{equation}
    (tan\varphi^D)^2+(tan\vartheta^D)^2=\frac{2}{\alpha}=-\frac{4E_k^S}{eU^*}\label{eq_Boundary}
\end{equation}

It indicates that, when the applied bias is larger than the critical value (\(U^*_C\)), the constant energy contours of the photoelectron cone boundary in the detector angle space are circle-like with its size determined by the kinetic energy and the corrected bias voltage as shown in \cref{eq_Boundary}. This is consistent with the measured results in Figs.~\ref{fig5_workfunction}a-e.

For the horizontal angle cut crossing the normal emission direction, where \(\vartheta^D=0\),  the direct relation between the binding energy and the detector angle can be obtained from \cref{eq_PhysicalE,eq_Boundary} as:
\begin{equation}
    E_B=h\nu-W_S-\frac{eU^*}{4}\cdot (tan\vartheta^D)^2\label{eq_BoundaryCurve}
\end{equation}
This equation describes a parabola-like relationship between the binding energy ($E_B$) and the detector angle ($\varphi^D$) for the given bias voltage. 

Figure~\ref{fig6_workfunctioncurve} presents the photoemission images in the detector angle measured along the horizontal angle cut crossing the normal emission for three typical materials: Bi\(_2\)Se\(_3\) (\cref{fig6_workfunctioncurve}a), Bi2212 (\cref{fig6_workfunctioncurve}b) and CsV\(_3\)Sb\(_5\) (\cref{fig6_workfunctioncurve}c). The corrected bias voltage applied in these measurements is all larger than the respective critical value (\(U^*_C\)). The corresponding curves from \cref{eq_BoundaryCurve} are plotted as dashed red lines in the figures. The boundary of the photoelectron cone is plotted as solid black lines. For Bi\(_2\)Se\(_3\) (\cref{fig6_workfunctioncurve}a), the measured boundary is in good agreement with that expected from \cref{eq_BoundaryCurve}. In contrast, the measured boundaries for Bi2212 (\cref{fig6_workfunctioncurve}b) and CsV\(_3\)Sb\(_5\) (\cref{fig6_workfunctioncurve}c) show clear deviations from the expected curves. To describe the measured boundaries, a correction term \(\eta\) needs to be introduced into \cref{eq_BoundaryCurve}, leading to the modified form:
\begin{equation}
    E_B=h\nu-W_S-\frac{eU^*\cdot \eta}{4}\cdot (tan\vartheta^D)^2
\end{equation}
Here, \(\eta\) represents the extent of deviation from the expected curve, with \(\eta=1\) corresponding to the perfect case. The correction terms required for Bi\(_2\)Se\(_3\) (\cref{fig6_workfunctioncurve}a), Bi2212 (\cref{fig6_workfunctioncurve}b) and CsV\(_3\)Sb\(_5\) (\cref{fig6_workfunctioncurve}c) are \(1.0\), \(1.18\) and \(1.27\), respectively. 

The observed boundary deviations indicate that, in some cases, the bias voltage experienced by the photoelectrons (effective bias voltage) may be different from the applied bias voltage. It means that, in these cases, the electric field that acts on the photoelectrons deviates from the ideal uniform field of the parallel plate capacitor assumption. This can arise from various sources, including the sample surface conditions, the beam size and the surrounding environments of the sample. This effect is not intrinsic to the material; it may vary between different measurements of the same material and even on different regions of the same sample surface. Therefore, the boundary measurement of the photoelectron cone as shown in \cref{fig6_workfunctioncurve} provides information of the local electric field to judge to what extent it is consistent with the ideal case of the parallel plate capacitor. Upon careful examination, we also found that the effective bias voltage can also be emission direction dependent, meaning that the correction term \(\eta\) can vary along different emission directions. Overall, these results demonstrate that while the parallel plate capacitance model captures the essential features of photoelectron emission, one should realize that the bias experienced by the photoelectrons may deviate from the applied voltage during  the bias ARPES measurements. The boundary analysis provides a practical way to check on the deviations. 

\section{Test of Bias ARPES and Validation of the Conversion Relationship}
\subsection{Feasibility Test of Bias ARPES Near the Normal Emission}
To test the performance of our bias ARPES system, we carried out ARPES measurements on Bi2212 by applying different sample bias voltages.
The Bi2212 sample is overdoped with a $T_c$ of 67\,K. The sample normal direction is aligned nearly along the analyzer lens axis. 
\cref{fig7_positionlimit}a shows the band structures of Bi2212 measured at 15\,K along the $\Gamma-Y$ direction using different sample bias voltages. In addition to the main Fermi surface (MR and ML in \cref{fig7_positionlimit}b), because  Bi2212 has superstructures along the $\Gamma-Y$ direction, it will form the first-order (L+1, L-1, R-1, R+1) and second-order (L+2, L-2, R-2, R+2) superstructure Fermi surface. It also has a shadow Fermi surface labeled as SB in \cref{fig7_positionlimit}b. Without the bias voltage, the observed bands consist of L-1, R-2, L-2 and R-1 in the topmost panel in \cref{fig7_positionlimit}a. When the bias voltage increases to 20\,V, the two main bands  (ML and MR) enter into the detector window. With further increase of the bias voltage to 90\,V, the SB, L+1, R+1 and SB bands come into the detector window (lowest panel in \cref{fig7_positionlimit}a). In this case, all the band structures expected from \cref{fig7_positionlimit}b are observed.  These clearly indicate that the collected emission angle increases with the increasing bias voltage.

Figure~\ref{fig7_positionlimit}c presents the photoemission images obtained from \cref{fig7_positionlimit}a by converting the detector angle into the sample emission angle. In the absence of the sample bias (U=0 V, top panel), the sample emission angle is identical to the detector angle. When the sample bias is applied (|U|=90 V), two different angle-conversion schemes can be tested. The middle panel of \cref{fig7_positionlimit}c shows the result obtained using the position limit (\cref{eq_Phi}), while the lower panel corresponds to the angular limit (\cref{eq_phiA}). \cref{fig7_positionlimit}d shows the corresponding angle distribution curves (ADCs) at the Fermi level extracted from \cref{fig7_positionlimit}c. A significant difference is observed between the two conversion schemes. Without bias voltage, two secondary superstructure bands (R-2 and L-2) are clearly observed (black curve in \cref{fig7_positionlimit}d). Their positions are intrinsic to the sample because no modification is involved in converting the detector angle to the emission angle. The peak position of the R-2 and L-2 bands produced from the position limit (red curve in \cref{fig7_positionlimit}d) matches closely to the standard positions without sample bias. On the other hand, the peak position of the two bands produced from the angular limit (blue curve in \cref{fig7_positionlimit}d) significantly deviates from the standard positions. These results indicate that, in our bias ARPES measurements, the position-limit is validated to describe the conversion between the detector angle and the emission angle. 

Figure~\ref{fig8_cut} shows the photoemission images measured under different bias voltages converted from the initial detector angle (\cref{fig8_cut}a) into the sample emission angle (\cref{fig8_cut}b) using the conversion relation in \cref{eq_DetectedE,eq_Phi} and further the electron momentum (\cref{fig8_cut}c) using \cref{eq_PhysicalE}. During the conversions, the spectral weight transformation is also considered by using \cref{eq_J,eq_Jacobian,eq_J1,eq_J2,eq_J3,eq_J4,eq_J5,eq_J6} and \cref{eq_JacobianMS}. When no bias is applied, the emission angle distribution at the Fermi level spans from -16$^\circ$ to 24$^\circ$ (leftmost panel in \cref{fig8_cut}b). As the bias increases, the emission angle range gradually expands, eventually approaching $\sim$90$^\circ$ at the bias of 90\,V (rightmost panel in \cref{fig8_cut}b), which corresponds to the maximum emission angle range for photoelectrons emitted from the surface.  \cref{fig8_cut}b clearly demonstrates that the collected emission angle is significantly expanded under bias, ultimately approaching $\pm90^\circ$ where all photoelectrons are detected by the analyzer. Correspondingly, the momentum range is also expanded with increasing bias as shown in \cref{fig8_cut}c. At the bias voltage of 90\,V, the maximum momentum reaches 1.01 $\pi/a$ as shown in the rightmost panel in \cref{fig8_cut}c. 

Figure~\ref{fig8_cut}d presents the angle distribution curves (ADCs) at the Fermi level, extracted from the photoemission images shown in \cref{fig8_cut}b. The detected emission angle range expands significantly with increasing bias, but the ADC peaks of the observed main bands (ML and MR) and superstructure bands all stay nearly at the same positions for different bias voltages. Figs.~\ref{fig8_cut}e-g provide a more quantitative analysis on the peak position of the ML (\cref{fig8_cut}e) and MR (\cref{fig8_cut}f) main bands, as well as their position difference (\cref{fig8_cut}g). Over the entire bias range, the ML peak stays at \(-(27.8\pm 0.6)^\circ\) while the MR peak stays at \((27.9\pm 0.3)^\circ\). Their position  difference remains essentially constant at \((55.8\pm 0.7)^\circ\). These results demonstrate the feasibility of increasing the emission angle range even up to the full 2\(\pi\) solid angle by applying the sample bias. They further confirm the internal consistency of the detector angle-emission angle conversion derived from the position limit. It was pointed out that the electric field from the hemispherical analyzer lens can cause distortions of the photoelectron trajectories\cite{majchrzak_2024_Rev.Sci.Instrum.}. The observed consistency across different bias voltages and the agreement with the position limit conversion (Fig.~\ref{fig8_cut}) suggest that such an effect is rather weak in our bias ARPES measurements. 

Now we move to test the feasibility of bias ARPES in two dimensional momentum space. \cref{fig9_FS}a shows Fermi surface mappings of Bi2212 in the detector angle space measured under different sample bias voltages using DA30 mode to deflect photoelectrons perpendicular to the analyzer slit direction. \cref{fig9_FS}b presents the corresponding conversion into the sample emission angle space ($\phi^S$ and $\theta^S$) obtained by using \cref{eq_DetectedE,eq_Phi,eq_theta}. \cref{fig9_FS}c presents the corresponding Fermi surface mappings obtained from \cref{fig9_FS}b by converting the sample emission angle into the electron momentum (k$_x$ and k$_y$) using \cref{eq_PhysicalE,eq_PhysicalKx,eq_PhysicalKy}. In these conversions, the spectral weight transformations are also considered by using \cref{eq_J,eq_Jacobian,eq_J1,eq_J2,eq_J3,eq_J4,eq_J5,eq_J6} and \cref{eq_JacobianMS}. As seen in \cref{fig9_FS}a, within the similar detector angle space, more and more Fermi surface features enter into the detection window with increasing bias voltage. After converting into the emission angle space (\cref{fig9_FS}b), it becomes clear that the collected emission angle expands significantly with increasing bias voltage, eventually approaching the full $\pm90^\circ$ range at a bias of 90\,V (rightmost panel in \cref{fig9_FS}b). Correspondingly, the momentum space coverage also expands with increasing bias voltage as shown in \cref{fig9_FS}c. At a bias of 90\,V, the maximum momentum reaches 1.01 $\pi/a$ as shown in the rightmost panel in \cref{fig9_FS}c, which is very close to the full 2\(\pi\) solid-angle limit for the 6.994 eV laser source. In the meantime, the overall Fermi surface topology stays at the same momentum position in all the measurements with different bias voltages. These results provide clear evidence that bias ARPES can be reliably extended to two-dimensional momentum space. The observed consistency across emission angle and momentum representations further supports the validity of the position limit used for the detector angle and the emission angle conversion.

\subsection{Effect of the Sample Bias on the Energy Resolution}

The high energy resolution is a core advantage of laser-based ARPES systems. It is necessary to investigate whether the sample bias application affects the energy resolution. To this end, we measured the Fermi edge of the polycrystalline gold at 11.7\,K by applying different bias voltages.

Figure~\ref{fig10_energyresolution}a displays a representative EDC acquired with a 2\,eV pass energy and a 0.1\,mm slit width, without applying the sample bias. The open circles are raw data; the black line is a Fermi-Dirac fit. The extracted Fermi-edge width (12-88\,\%) is 4.09\,meV which is consistent with all the contributions from the thermal broadening (11.7\,K corresponds to 4.04\,meV), the corresponding energy resolution of the analyzer (0.5\,meV) and the linewidth of the laser (0.26\,meV). This gives an instrumental energy resolution better than 1\,meV. We measured the gold Fermi edge at 11.7\,K by using 2\,eV, 5\,eV and 10\,eV pass energies at different sample bias voltages. \cref{fig10_energyresolution}b shows a series of typical EDCs measured using 10\,eV pass energy and 0.1\,mm slit width under different bias voltages. These EDCs are fitted by the Fermi-Dirac distribution function and the extracted Fermi level width is plotted in \cref{fig10_energyresolution}c as a function of the sample bias voltage (blue circles in \cref{fig10_energyresolution}c). \cref{fig10_energyresolution}c also includes the extracted Fermi-edge widths measured by using 5\,eV pass energy and 0.1\,mm slit width (red circles), as well as 2\,eV pass energy and 0.1\,mm slit width (black circles). 

The analyzer energy resolution for 2\,eV, 5\,eV and 10\,eV pass energies with a 0.1\,mm slit width is 0.50\,meV, 1.25\,meV and 2.50\,meV, respectively. These give the overall instrumental energy resolution of 0.56\,meV, 1.28\,meV and 2.51\,meV for the 2\,eV, 5\,eV and 10\,eV pass energies, respectively, after considering the laser linewidth of 0.26\,meV. The corresponding expected Fermi edge widths at 11.7\,K are 4.08\,meV (black solid line in \cref{fig10_energyresolution}c), 4.23\,meV (red solid line) and 4.76\,meV (blue solid line), respectively, after further considering the thermal broadening. The measured Fermi edge width at zero bias voltage is 4.09\,meV, 4.46\,meV and 5.12\,meV for the 2\,eV, 5\,eV and 10\,eV pass energies, respectively, giving an overall instrumental energy resolution of 0.67\,meV, 1.89\,meV and 3.15\,meV, respectively, after removing the contribution from the thermal broadening. 

As seen in \cref{fig10_energyresolution}c, the measured Fermi level width exhibits a slight increase with the increase of the bias voltage for all the three pass energy cases. For the 2\,eV pass energy, the measured width increases from 4.09\,meV at zero bias to 4.97\,meV at 18\,V corresponding to the measured energy resolution change from 0.67\,meV to 2.90\,meV. For the 5\,eV pass energy, the measured width increases from 4.46\,meV at zero bias to 5.48\,meV at 49\,V corresponding to the measured energy resolution variation from 1.89\,meV to 3.65\,meV. For the 10\,eV pass energy, the measured width increases from 5.12\,meV at zero bias to 6.45\,meV at 99\,V corresponding to the measured energy resolution degradation from 3.15\,meV to 5.03\,meV. These results indicate that the sample bias application can cause extra energy broadening. This is likely related to the stability level of the bias voltage that the source meter can provide. These results also demonstrate that the energy resolution of our bias ARPES system remains consistently high even in the presence of applied sample bias and it can still reach \(\sim\)5\,meV even using 10\,eV pass energy and 100\,V bias voltage. Future efforts to minimize the effect from the source meter will likely improve the energy resolution even further.

\subsection{Effect of the Sample Bias on the Angular Resolution}

The angular resolution and the corresponding momentum resolution represent other key performance parameters of ARPES systems. To test the effect of the sample bias application on the angular resolution, we performed ARPES measurements on the optimally-doped Bi2212 (\(T_c=91\,K\)) at 17\,K along the $\Gamma-Y$ direction (similar cut in \cref{fig7_positionlimit}b) under different sample bias voltages by using the 30-degree angular mode. \cref{fig11_angularresolution}a shows the ADCs at the Fermi level plotted as a function of the detector angle for different bias voltages ranging from 0 to 99\,V. After converting the detector angle into the sample emission angle, \cref{fig11_angularresolution}b presents the corresponding ADCs as a function of the emission angle. Here we will focus on the two ADC peaks of the main bands (ML and MR) as marked in \cref{fig11_angularresolution}a and \ref{fig11_angularresolution}b. In the detector angle space (\cref{fig11_angularresolution}a), the two peaks move closer with increasing bias voltage. In contrast, in the emission angle space (\cref{fig11_angularresolution}b), the two peaks remain nearly at the same positions for all bias voltages. From the position of the two peaks, considering the sample work function of 4.154\,eV and the laser photon energy of 6.994\,eV, we can calculate the angular magnification factor ($M_A$) of the main bands (ML and MR) as a function of the bias voltage using \cref{eq_Jacobian,eq_J1,eq_J2,eq_J3,eq_J4,eq_J5,eq_J6}. The results are plotted in \cref{fig11_angularresolution}c. The full width at half maximum (FWHM) of the two ADC peaks in the detector angle space (\cref{fig11_angularresolution}a) is extracted and plotted in \cref{fig11_angularresolution}d as a function of the bias voltage (open diamonds for ML band and filled diamonds for MR band). \cref{fig11_angularresolution}e shows the extracted ADC width (FWHM) of the ML (open circles) and MR (filled circles) bands in the emission angle space obtained from \cref{fig11_angularresolution}b. In the detector angle space, the ADC width of the two main bands (ML and MR) shows a slight increase with increasing bias voltage (\cref{fig11_angularresolution}d) from \(\sim\)0.29\(^\circ\) at 20\,V to \(\sim\)0.41\(^\circ\) at 99\,V. In contrast, in the emission angle space, the ADC width of the two main bands exhibits an obvious increase with increasing bias voltage (\cref{fig11_angularresolution}e), from \(\sim\)0.70\(^\circ\) at 20\,V to \(\sim\)1.60\(^\circ\) at 99\,V. 

In the ideal model of parallel plate capacitance, the observed angular width of a given band is determined by its intrinsic angular width of the sample (\(dA^S\)), the angular resolution of the analyzer (\(dA^D\)) and the corresponding angular magnification factor (\(M_A\)). The expected angular width in the detector angle space (\(dA^{Ex}_D\))  can be expressed as:
\begin{equation}
    dA^{Ex}_D=\sqrt{(dA^S/M_A)^2+(dA^D)^2}\label{eq_AngularResolution}
\end{equation}
while the expected angular width in the emission angle space (\(dA^{Ex}_E\)) can be expressed as:
\begin{equation}
    dA^{Ex}_E=\sqrt{(dA^S)^2+(M_A\times dA^D)^2}\label{eq_expectedAngular}
\end{equation}

At the bias voltage of 20\,V, we observed a narrow ADC peak width of \(0.29^\circ\) for the ML band in the detector angle space as shown in \cref{fig11_angularresolution}f. Also from the variation of the ADC peak width as a function of the bias voltage in the emission angle space as shown in \cref{fig11_angularresolution}e, the peak width at zero bias is estimated to be \(0.50^{\circ}\). From these results and considering \cref{eq_AngularResolution,eq_expectedAngular}, we  get the approximate value of the intrinsic angular width of the sample (\(dA^S\approx 0.46^\circ\)) and the upper limit of the angular resolution of the analyzer (\(dA^D\approx 0.19^\circ\)). Using these values, we calculate the expected angular widths in both the detector angle space and the emission angle space using \cref{eq_AngularResolution,eq_expectedAngular}. The calculated results are plotted as dashed lines in \cref{fig11_angularresolution}d and \cref{fig11_angularresolution}e, respectively.

In \cref{fig11_angularresolution}d, the measured ADC width of the two main bands in the detector angle space deviates from the expected trend and the difference becomes larger with increasing bias voltages. In \cref{fig11_angularresolution}e, the measured ADC width of the two main bands (ML and MR) in the emission angle space also deviates from the expected curve and the difference gets larger with the increasing bias voltage. These results indicate that the application of the sample bias gives rise to extra peak broadening that degrades the angular resolution. Possible sources of this deviation include non-uniformities in the applied electric field and the finite laser spot size. These measurements provide an upper bound for the angular resolution in the detector angle space which is better than \(0.19^\circ\). We also note that, when the angular magnification factor is high due to high sample bias or large emission angle, the angular resolution in the emission angle space is then dominated by \(dA^D\times M_A\). These measurements and analyses provide important information to achieve a high measured angular resolution in the emission angle space in the bias ARPES experiments. 

\subsection{Effect of the Beam Size on the Bias ARPES Measurement}

The size of the laser spot on the sample surface is a critical parameter for bias ARPES measurements. To test its effect, we carried out a series of measurements on optimally-doped Bi2212 (\(T_c=91\,K\)) at 17\,K along the \(\Gamma-Y\) direction by varying the laser spot from \(20\times 20\,\mu m^2\) to \(240\times 104\,\mu m^2\). Here the size of the elliptical laser spot is defined as \(\sigma_x\times \sigma_z\) where \(\sigma_x\) represents the spot size along the horizontal direction while \(\sigma_z\) represents the size along the vertical direction. The effective spot size can be defined as \(\sqrt{\sigma_x\times \sigma_z}\). \cref{fig12_beamsize}a displays the photoemission images in the analyzer detector angle space measured at different spot sizes under a bias voltage of 90\,V. The corresponding images in the sample emission angle space, converted from \cref{fig12_beamsize}a, are shown in \cref{fig12_beamsize}b. In this case, many features can be covered in the angle window, including  the main bands (ML and MR), first-order superstructure bands (\(L\pm1,~ R\pm1\)) and second-order superstructure bands (\(L\pm2,~ R\pm2\)), as indicated in \cref{fig12_beamsize}a and \cref{fig12_beamsize}b. For comparison, we also carried out zero bias measurements using the same set of spot sizes as shown in \cref{fig12_beamsize}c. In this case, only the two second-order superstructure bands (R-2 and L-2) and the two first-order superstructure bands (L-1 and R-1) are covered in the angle window.

As seen in Figs.~\ref{fig12_beamsize}a-b with the bias voltage of 90\,V applied, when the spot size is small (\(20\times 20\,\mu m^2\)), the observed bands are sharp and well defined. As the spot size increases, the bands gradually become broader. When the spot size reaches \(240\times 104\,\mu m^2\), the bands get seriously blurred, and several features (L-1, R-2, L-2 and R-1) become invisible. In contrast, without sample bias, the spot size has much weaker effect on the observed band structures (L-1, R-2, L-2 and R-1) as seen in \cref{fig12_beamsize}c. To quantitatively analyze the effect, \cref{fig12_beamsize}d shows the ADCs at the Fermi level extracted from \cref{fig12_beamsize}b. The ADCs at the Fermi level obtained from \cref{fig12_beamsize}c are shown in \cref{fig12_beamsize}e. The corresponding peaks stay at similar positions, indicating that the variation of the spot size does not affect the conversion relation between the detector angle and the emission angle. The ADC peak width of the ML and MR main bands in \cref{fig12_beamsize}d is extracted and plotted in \cref{fig12_beamsize}f (red circles). The ADC peak width of the R-2 and L-2 superstructure bands in \cref{fig12_beamsize}e is obtained and shown in \cref{fig12_beamsize}f (blue circles). With the application of 90\,V bias voltage, the emission ADC width of the two main bands (ML and MR) increases significantly with the increasing spot size, from \(\sim\)\(1.9^\circ\) at the spot size of \(20\times 20\,\mu m^2\) (corresponding effective spot size of \(20.0\mu m\)) to \(\sim\)\(8.7^\circ\) at the spot size of \(240\times 104\,\mu m^2\) (corresponding effective spot size of \(158.0\mu m\)). In contrast, without the sample bias,  the two second-order superstructure bands (R-2 and L-2) broaden only slightly, from \(\sim\)\(1.5^\circ\) to \(\sim\)\(2.4^\circ\). These results demonstrate that the beam  size has a profound effect on the performance of the bias ARPES measurements and small laser beam is essential to yield sharp features and high instrumental resolution. 

\subsection{Feasibility Test of Bias ARPES Off the Normal Emission}

As we have shown before, the bias ARPES can significantly expand the covered momentum space. But in the meantime, the application of the sample bias also degrades the energy and angular resolutions. In particular, the degradation of the angular resolution is directly related to the angular magnification factor (\(M_A\)). In fact, by symmetry considerations, it is not necessary to cover all the four quadrants in the Brillouin zone. Instead, depending on the symmetry, it is sufficient to cover part of the Brillouin zone to extract related information. For example, for a material with four-fold symmetry, covering half of the first quadrant is enough because the eight parts of the Brillouin zone are equivalent. The coverage of only a quadrant or part of a quadrant asks ARPES to measure off the sample normal emission. It is then natural to ask whether the bias ARPES can still work well when the sample is tilted away from the normal emission direction.

To test this, we carried out two kinds of bias ARPES measurements on optimally-doped Bi2212 (\(T_c=91\,K\)) at 17\,K. One is to use a high bias voltage (\(|U|=80\,V\)) to cover a large momentum space and then tilt the sample along the \(\Gamma-Y\) direction with different angles from \(\Phi = 1.3^\circ\) (normal emission) to \(\Phi = 16.7^\circ\). The measured Fermi surface mappings are shown in \cref{fig13_rotation}a. The other way is to tilt the sample with different angles from \(\Phi = 0.8^\circ\) to \(\Phi = 31.2^\circ\) and in the meantime gradually reducing the bias voltage from 99\,V to 0\,V. The measured Fermi surface mappings are presented in \cref{fig13_rotation}b. In both cases, the covered momentum space shrinks and shifts into the first quadrant. 

As seen in \cref{fig13_rotation}a and \ref{fig13_rotation}b, when the sample is tilted more and more off the normal emission direction, the observed Fermi surface topology remains similar with all the observed Fermi surface sheets staying at similar positions in the momentum space. To further check on this more quantitatively, \cref{fig13_rotation}c shows the momentum distribution curves (MDCs) at the Fermi level along the \(\Gamma-Y\) direction extracted from \cref{fig13_rotation}a. \cref{fig13_rotation}d presents the MDCs at the Fermi level along the \(\Gamma-Y\) direction extracted from \cref{fig13_rotation}b. In both cases, the peaks of the ML and MR main bands and the R+1 superstructure band stay at similar positions within the experimental uncertainty. These results demonstrate clearly that the bias ARPES can still work well when the sample is tilted off the normal emission direction. Note that, the shrinkage of the covered momentum space from four quadrants to one quadrant or part of one quadrant dramatically reduces the sample bias voltage from 99\,V to \(\sim\)30\,V. This in turn significantly improves the energy and angular resolutions. 

\section{Demonstration of Full $2\pi$ Solid Angle Collection in Laser ARPES Measurements}

\subsection{Case Study 1: Bi2212 }

Bi2212 is a prototypical high temperature cuprate superconductor which is most extensively studied by ARPES measurements. Laser ARPES, with its super high energy resolution and high data statistics, has gained new insights by measuring Bi2212 using 6.994\,eV laser source\cite{zhang_2008_Phys.Rev.Lett.a,zhang_2008_Phys.Rev.Lett.,bok_2016_Sci.Adv.,yan_2023_Proc.Natl.Acad.Sci.U.S.A.,gao_2024_NatCommun,zhou_2018_Rep.Prog.Phys.}. However, due to the limitation of the relatively low photon energy and the mechanical constraints on the accessible emission angles, the previous laser APRES measurements can cover only a small part of the first Brillouin zone. To date, laser ARPES measurements with 6.994\,eV laser have not fully reached the $(\pi,0)$ antinodal point, which is pivotal for unraveling the mechanisms underlying high temperature superconductivity in cuprates. 

By using bias ARPES, for the first time, we can collect the full $2\pi$ solid angle of photoelectrons and reach the $(\pm \pi,0)$ point and $(0,\pm \pi)$ point in Bi2212 in laser ARPES measurements using 6.994\,eV laser. Figure \ref{fig14_Bi2212} presents a comprehensive full two-dimensional Fermi surface mappings for Bi2212 measured under four different polarization geometries. The measured Bi2212 sample is overdoped with a $T_c$ of 67\,K. We first measured the work function of the sample and it is 4.377\,eV. The horizontal detector angle of our DA30L analyzer ($\varphi^D$) covers $\pm19.6^\circ$. The DA30 mode can deflect the vertical detector angle between $\pm15^\circ$. To collect 2\(\pi\) solid angle along all the directions, in particular along the vertical direction, the minimum sample bias required is 147.3\,V according to \cref{fig3_simulation}. The measurements in \cref{fig14_Bi2212} were performed with a sample bias of 140\,V using 20\,eV pass energy and 0.1\,mm slit width. These Fermi surface mappings are obtained by scanning the vertical detector angle every 0.1$^\circ$ without shifting or rotating the sample. In the sample work function measurement of this particular sample, we get a correction term \(\eta =1.18\) to fit the photoelectron cone boundary. This indicates that the effective bias voltage experienced by photoelectrons is \(\eta \cdot U^* = 161 \,V\) as shown in \cref{fig6_workfunctioncurve}. This effective bias voltage is used in converting the detector angle to the emission angle and momentum in \cref{fig14_Bi2212}. In \cref{fig14_Bi2212}a, the Fermi surface mapping was measured by aligning the Cu-O-Cu direction (\(\Gamma-M\)) of the sample along the slit while setting the electric vector of the laser light perpendicular to the slit, as marked in the bottom-left corner. In \cref{fig14_Bi2212}b, the electric vector is switched to be parallel to the slit. In \cref{fig14_Bi2212}c and 14d, the diagonal direction (\(\Gamma-Y\)) of Bi2212 is aligned along the slit while the electric vector of laser light is set perpendicular (\cref{fig14_Bi2212}c) or parallel (\cref{fig14_Bi2212}d) to the slit. The nodal Fermi momentum of the main ML and MR bands obtained from Figs.~\ref{fig14_Bi2212}a-d is \(\sim\)0.51\,$\pi/a$ for this overdoped Bi2212 sample. This is consistent with the value reported before\cite{zhang_2016_SB}.

In \cref{fig14_Bi2212}a, the red circle indicates the maximum momentum space that can be accessed when the full $2\pi$ solid angle of photoelectrons is collected in our laser ARPES measurements using 6.994\,eV laser. It is clear from the Fermi surface mappings in Figs.~\ref{fig14_Bi2212}a-d that the application of a sample bias allows the analyzer to collect photoelectrons over the entire $2\pi$ solid angle, as evidenced by the covered momentum space approaching the red circle in \cref{fig14_Bi2212}a. For the first time, it is possible to measure ($\pm \pi$,0) and (0,$\pm \pi$) antinodal points in laser ARPES measurements using 6.994\,eV laser. We note that, the work function of Bi2212 decreases with the decreasing of the doping level (Table~1). Since we can already cover the ($\pi$,0) point in the overdoped Bi2212, it is possible to cover even larger momentum space in the optimally doped and underdoped Bi2212 samples.

Our bias ARPES measurements also allow, for the first time, to study the photoemission matrix element effects in laser ARPES measurements of Bi2212 using 6.994\,eV laser. The measured data in Figs.~\ref{fig14_Bi2212}a-d show several advantages in matrix element analysis. First, all the bands, including the main bands, first-order superstructure bands, second-order superstructure bands and the shadow bands, are observed. Second, the polarization geometry keeps fixed because the sample and the laser light do not change during the measurements. Third, Figs.~\ref{fig14_Bi2212}a-d have covered all possible linear polarization geometries by combining the sample orientation and the light polarization. 

The Fermi surface mappings in Figs.~\ref{fig14_Bi2212}a-d reveal strong matrix element effects induced by different linear polarization geometries. Here we focus on the main Fermi surface in the first (MR) and third (ML) quadrants because the main Fermi surface in the second and fourth quadrants (MU and MD) are complicated by the superstructure Fermi surface. The main Fermi surface consists of bonding (BB)  and antibonding (AB) Fermi surface sheets due to bilayer splitting in Bi2212. Notably, while the main Fermi surface intensity is similar in all the four quadrants in \cref{fig14_Bi2212}a and 14b, it shows a dramatic difference between (1,3) quadrants and (2,4) quadrants in \cref{fig14_Bi2212}c and \ref{fig14_Bi2212}d. The main Fermi surface (MR and ML) are strong in intensity when measured in the first three polarization geometries (Figs.~\ref{fig14_Bi2212}a-c) but becomes very weak in intensity when measured in the fourth polarization geometry (\cref{fig14_Bi2212}d). The bilayer splitting is clearly resolved near \((\pm \pi,0)\) but mainly the antibonding Fermi surface is visible near the $(0,\pm \pi)$ region as seen in \cref{fig14_Bi2212}a. But it is reversed in \cref{fig14_Bi2212}b where the bilayer splitting becomes clear near the $(0,\pm \pi)$ but mainly the antibonding Fermi surface is visible near the $(\pm \pi,0)$ region. In \cref{fig14_Bi2212}c, mainly the antibonding Fermi surface is observed and the bonding Fermi surface is rather weak. The strong matrix element effects are closely related to the orbital characters of the main Fermi surface which consist of the Cu $d_{x^2 - y^2}$ and O 2$p_{x/y}$ orbitals. Such detailed momentum and polarization dependence of the matrix element effects in Bi2212 can be used to guide ARPES measurements.

Figures~\ref{fig14_Bi2212}e-h display the band structures measured along the $\Gamma-M$ momentum cut (marked as Cut1 in \cref{fig14_Bi2212}d) obtained from Figs.~\ref{fig14_Bi2212}a-d, respectively. Figs.~\ref{fig14_Bi2212}i-m show the band structures measured along the $\Gamma-Y$ momentum cut (marked as Cut2 in \cref{fig14_Bi2212}d) obtained from Figs.~\ref{fig14_Bi2212}a-d, respectively. Here, our measured energy range extends to the binding energy of 1.4\,eV. Since the momentum cuts along the vertical detector angle $\vartheta^D$ are densely taken (every 0.1\,$^\circ$ in the $\pm15^\circ$ range), after interpolation, it allows us to obtain high quality images of band structures along any momentum cut. All the expected features, including the main bands, superstructure bands, shadow band and the high energy waterfall features, can be identified in these laser-based bias ARPES measurements. Their intensity also shows  strong matrix element effects. These results measured under different polarizations provide more information on studying the high energy nodal dispersion\cite{zhang_2008_Phys.Rev.Lett.a}, the antinodal von Hove singularity and the high energy waterfall features\cite{zhang_2008_Phys.Rev.Lett.a} in Bi2212. 

Since Bi2212 has a superstructure with the wavevector along the \(\Gamma-Y\) direction, by symmetry analysis, it is enough to cover the \(\Gamma-Y-X-\Gamma\) region because the first Brillouin zone consists of four equivalent such regions. In fact, if we focus on the main Fermi surface MR, it is sufficient to cover half of the first quadrant, i.e., the \(\Gamma-M-Y-\Gamma\) region. \cref{fig16_offnormalEp5} shows the measured Fermi surface mapping of Bi2212 (optimally-doped, \(T_c=91\,K\)) at 17\,K. Here the sample was tilted along \(\Gamma-M\) by 11.5 degree and the applied sample bias is 40\,V, much lower than the 140\,V used in \cref{fig14_Bi2212}. This allows us to use 5\,eV pass energy which gives an analyzer energy resolution of 1.25\,meV for the 0.1\,mm slit. The lower sample bias of 40\,V also gives a lower angular magnification factor (see \cref{fig4_magnification}) which in turn leads to a better momentum resolution. As seen in \cref{fig16_offnormalEp5}, the main Fermi surface in the \(\Gamma-Y-X-\Gamma\) region is fully covered and the covered momentum space reaches 1.04\(\pi/a\) near the M point. The data contains all the information about the main Fermi surface and the main band. These results demonstrate that these off-normal bias ARPES measurements can efficiently cover the desired momentum region with lower bias voltages, thus preserving superior energy and angular resolutions. This approach is particularly advantageous for materials with high symmetry, where only a fraction of the Brillouin zone needs to be measured. Therefore, bias ARPES not only enables full 2\(\pi\) solid angle collection but also offers flexible measurement strategies to optimize both momentum coverage and instrumental performance.

\subsection{Case Study 2: CsV$_3$Sb$_5$}

As a second showcase of the $2\pi$ solid-angle collection, we carried out bias ARPES measurements on CsV$_3$Sb$_5$, a new kagome superconductor\cite{ortiz_2019_Phys.Rev.Mater.,wilson_2024_NatRevMater,yang_2025_ChinesePhys.B}. Since its discovery\cite{ortiz_2019_Phys.Rev.Mater.}, extensive efforts have been devoted to study its electronic structure. Here we present high resolution measurements acquired with laser ARPES using a 6.994\,eV laser source. By applying a sample bias, the full 2\(\pi\) solid angle of photoelectrons is collected across all emission directions and the accessed momentum space fully covers the first Brillouin zone and even reaches large part of  the second Brillouin zone.

Figures~\ref{fig15_CVS}a-d display the measured constant energy contours of CsV$_3$Sb$_5$ at different binding energies of 0\,eV (Fermi surface, \cref{fig15_CVS}a), 0.1\,eV (\cref{fig15_CVS}b), 0.3\,eV(\cref{fig15_CVS}c), and 0.6\,eV (\cref{fig15_CVS}d). The sample work function was measured to be 1.345\,eV. The data were collected at 94\,K with a sample bias of 200\,V, using a 20\,eV pass energy and a 0.1\,mm slit width. To minimize the effect of the polarization selection rules and enhance overall band observation, circularly polarized light was used. In the sample work function measurement of this particular sample, we get a correction term \(\eta =1.27\) to fit the photoelectron cone boundary as shown in \cref{fig6_workfunctioncurve}. The effective bias voltage (\(\eta \cdot U^* = 250\,V \)) is used in converting the detector angle to the emission angle and momentum in \cref{fig15_CVS}. We note that, the effective bias voltage in  this case is much higher than the applied bias voltage. This is possibly related to the surface condition of the sample and the exact reason needs further investigation. The measured Fermi surface (\cref{fig15_CVS}a) and constant energy contours (Figs.~\ref{fig15_CVS}b-d) are consistent with the  symmetry of the material and, in particular, agrees well with the corresponding BZ boundaries (blue hexagon for the first BZ in Figs.~\ref{fig15_CVS}a-d and black hexagons for the second BZ in \cref{fig15_CVS}a). They are also consistent with the previous ARPES measurements on AV$_3$Sb$_5$ (A= K, Rb and Cs)\cite{luo_2022_NatCommuna,hu_2022_NatCommun}. These results indicate that, to get the bias ARPES work properly, it is important to realize there is a correction term \(\eta\) to the applied bias voltage and it is necessary to determine the effective bias voltage (\(\eta \cdot U^* \)) correctly. 

In \cref{fig15_CVS}a, the red circle represents the maximum momentum space that can be measured when all the \(2\pi\) solid angle of photoelectrons is collected using 6.994\,eV laser. It is clear that our covered momentum space fully approaches the red circle, demonstrating that all  the full \(2\pi\) solid angle of photoelectrons across all emission directions are collected. Since CsV$_3$Sb$_5$ has a low sample work function, the accessed momentum space completely covers the first BZ (blue hexagon in Figs.~\ref{fig15_CVS}a-d) and extends well toward the center of the second BZ (black hexagons in \cref{fig15_CVS}a). The electron-like \(\alpha\) Fermi surface around the zone center, the hexagonal  hole-like \(\beta\) Fermi surface, the triangular hole-like  \(\gamma\) Fermi surface around K and the triangular electron-like \(\delta\) Fermi surface around K are all clearly observed. The \(\gamma\) and \(\delta\) Fermi surface are nearly degenerate in \cref{fig15_CVS}a but they become separated in the constant energy contour of the binding energy 0.1\,eV in \cref{fig15_CVS}b. In particular, a tiny electron-like pocket (\(\epsilon\)) around the M point is  clearly resolved in \cref{fig15_CVS}a. The measured Fermi surface topology are consistent with the previous ARPES measurements and agrees fully with the band structure calculations\cite{ortiz_2019_Phys.Rev.Mater.,luo_2022_NatCommuna,hu_2022_NatCommun}.  

Figure~\ref{fig15_CVS}e presents band structures of  CsV$_3$Sb$_5$ extracted along representative high symmetry momentum cuts marked in \cref{fig15_CVS}d. The bias ARPES experiments make it possible to measure band structures along all the high symmetry momentum cuts even using the laser source of 6.994\,eV. The dense angular sampling along the vertical analyzer angle direction (every 0.1\(^\circ\) between \(\pm15^\circ\)) allows us to get high quality band structures along any arbitrary momentum cuts. The \(\alpha\), \(\beta\), \(\gamma\), \(\delta\) and especially \(\epsilon\) bands are all clearly observed. The hallmark spectral features of CsV$_3$Sb$_5$ are also resolved, including the Dirac point (DP) at K and the von Hove singularities (vHs1, vHs2 and vHs3) at M. The bias ARPES has solved the long-standing problem of limited momentum space coverage in laser ARPES measurements.  But it keeps all the advantages of laser ARPES, including super high energy resolution, high momentum resolution and high data statistics. These make it possible to carry out further ARPES studies on the charge density wave (CDW), superconductivity and many-body interactions in CsV$_3$Sb$_5$.

\section{Discussion and Conclusion}

In this work, we have demonstrated that bias ARPES can be successfully implemented in our laser ARPES system. By applying a sample bias, the accessible momentum space is greatly expanded, ultimately allowing full 2\(\pi\) solid angle collection of photoelectrons. This capability is realized by minimal modification of our existing laser ARPES system by only inserting a thin sapphire piece to electrically isolate the sample. In doing so, it has addressed a long-standing drawback of laser ARPES that can cover only a limited momentum space. In the meantime, it keeps all of the unique advantages of laser ARPES, including the ultrahigh energy resolution, fine momentum resolution, low sample temperature, enhanced bulk sensitivity and high data statistics. 

We carefully studied the converting relations between the detector angle, the emission angle and the electron momentum in bias ARPES measurements. We demonstrated that the parallel-plate capacitor model works well in our bias ARPES measurements. In addition, the position limit provides a proper description for converting the detector angle into the emission angle. A precise approach is developed to determine the sample work function which is critical in the angle-momentum conversion of the bias ARPES experiments. 
We further investigated the performance of bias ARPES in detail.  The application of a bias voltage leads to a minor degradation of the energy resolution, primarily due to the voltage fluctuation from the source meter, but it keeps the high energy resolution (better than 5\,meV) even with the application of a high bias voltage (\(\sim\)100\,V). The angular resolution is also degraded by the sample bias mainly because of the angular magnification and an additional contribution that is dependent on the beam size. We found that the beam size plays a critical role on the performance of bias ARPES and a small beam size is essential for achieving sharp features and high instrumental resolution. We further demonstrated that bias ARPES can still work when the sample is tilted off the normal emission direction. This makes it possible to cover one quadrant or part of one quadrant of the Brillouin zone, which significantly reduces the required sample bias voltage and thus improves both the energy and angular resolutions.

\begin{figure*}[tbp]
    \begin{center}
    \includegraphics[width=1.0\columnwidth,angle=0]{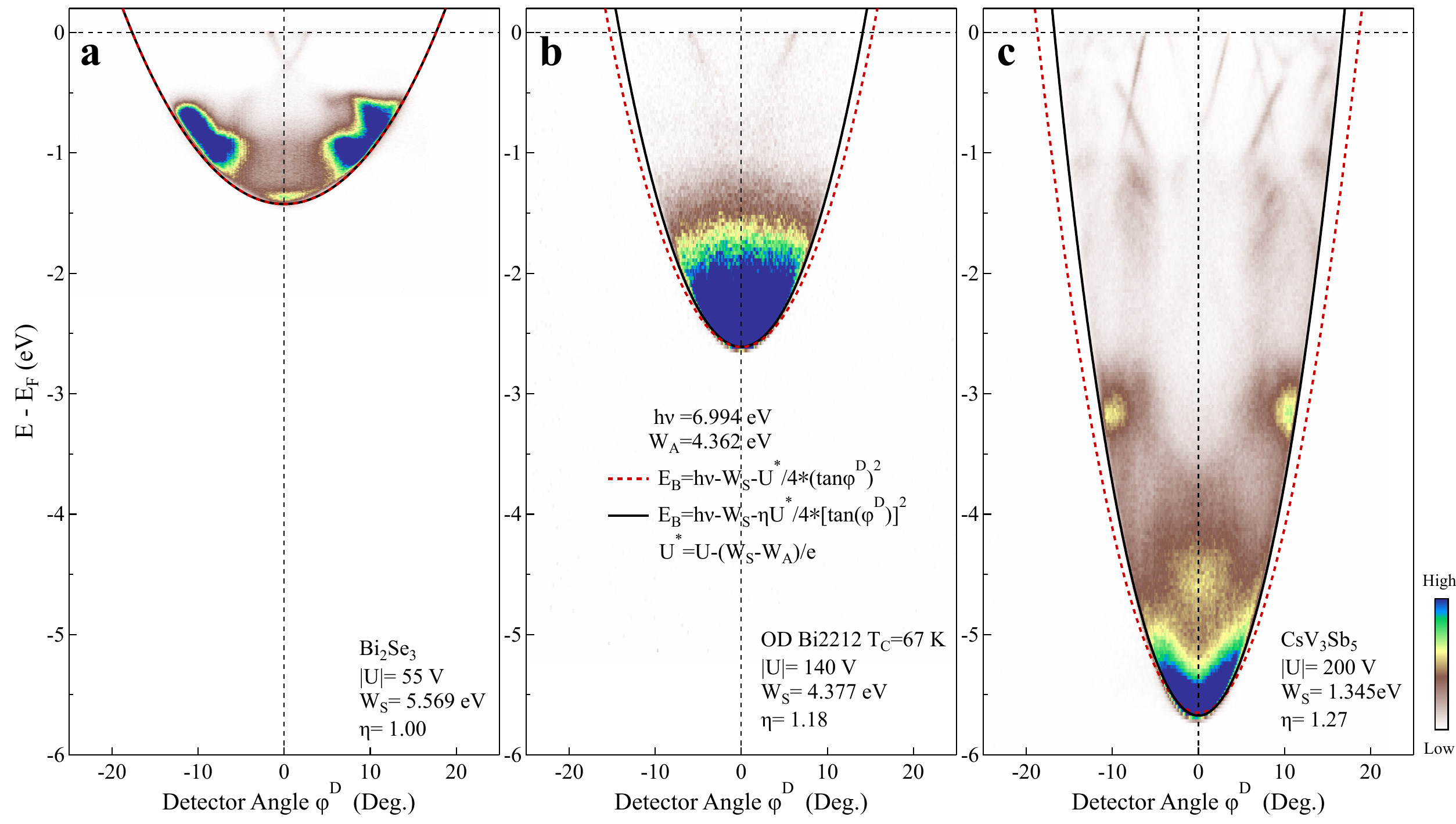}
    \end{center}
    \caption{\textbf{Photoemission Images along a Horizontal Angle Cut Crossing the Normal Emission Direction and Description of the Boundaries.} At the sufficiently high bias voltage, the boundary in the energy-detector angle space corresponds to photoelectrons with the emission angle of 2\(\pi\) at different binding energies. In the perfect environment of parallel plate capacitor and in the position limit, the boundary can be described by the relation \(E_B=h\nu-W_S-U^*/4\times (tan\varphi^D)^2\) (red dashed lines), where \(U^*=[U-(W_A-W_S)/e]\). In practice, the measured boundary may deviate from the ideal case and a correction term \(\eta\) needs to be introduced: \(E_B=h\nu-W_S-U^*/4\times [tan(\varphi^D\cdot \eta)]^2\) (black solid lines). {\textbf{(a,)}} Photoemission image of Bi\(_2\)Se\(_3\) measured at a sample bias voltage of 55\,V. Here the boundary can be well described by the ideal case with \(\eta=1.0\). {\textbf{(b,)}} Photoemission image of Bi2212 measured at a sample bias voltage of 140\,V. Here the boundary slightly deviates from the ideal case with the correction term \(\eta=1.18\).{\textbf{(c,)}} Photoemission image of CsV\(_3\)Sb\(_5\) measured at a sample bias voltage of 200\,V. Here the boundary deviates from the ideal case with the correction term \(\eta=1.27\).
    }
    \label{fig6_workfunctioncurve}
\end{figure*}

\begin{figure*}[tbp]
\begin{center}
\includegraphics[width=0.85\columnwidth,angle=0]{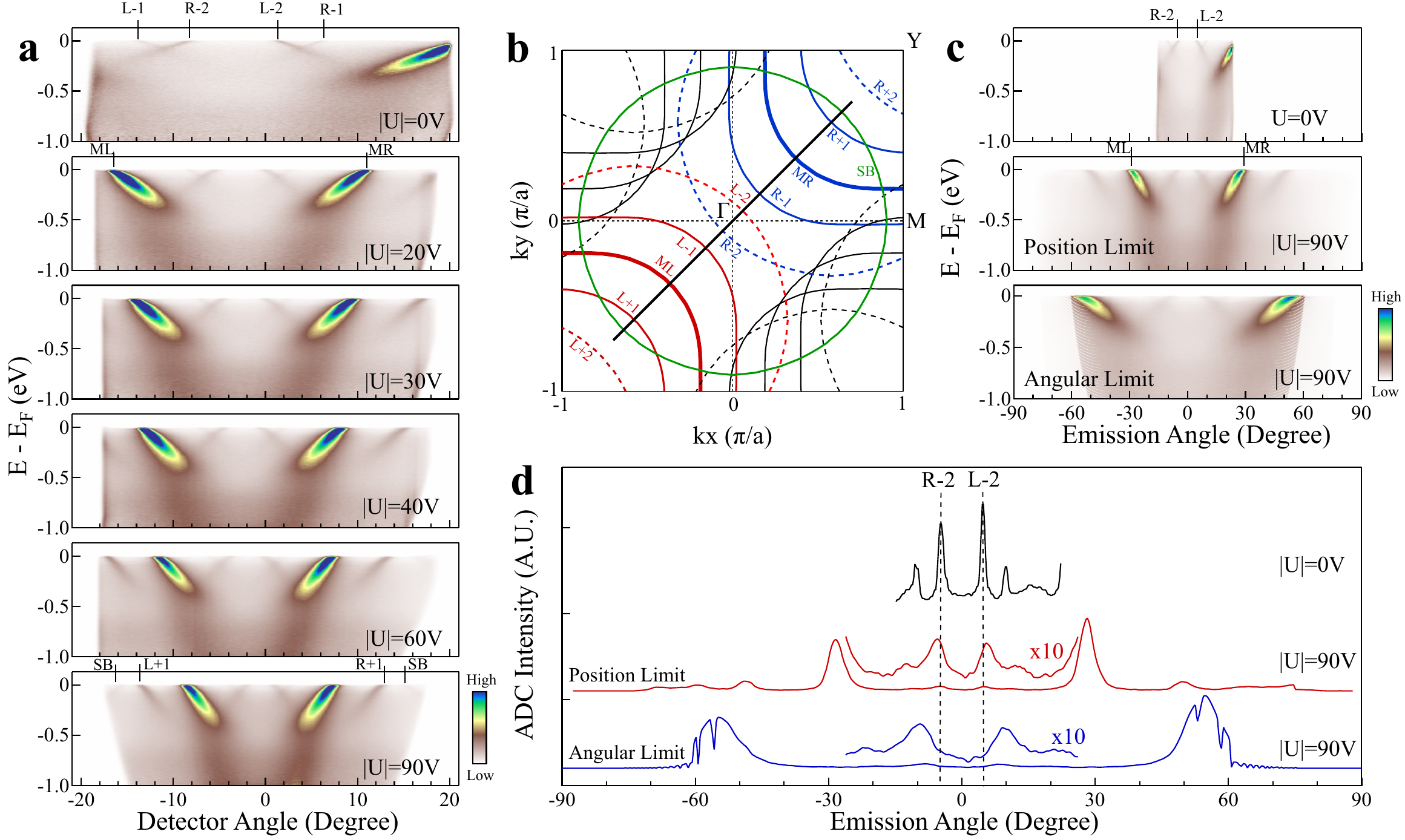}
\end{center}
\caption{\textbf{Band Structures Measured under Different Sample Bias Voltages and Validation of the Position Limit.}~{\textbf{a,}} Original photoemission images on the analyzer detector measured on Bi2212 with different sample bias voltages. The 30-degree angular mode of our DA30L analyzer can cover [-19.6,19.6] angle range at zero bias (topmost image). The direction of the corresponding momentum cut is shown by the black line in \textbf{b}. The measured images consist of the strong main bands (marked as ML and MR), weak first-order superstructure bands (marked as L+1, L-1, R+1 and R-1), second-order superstructure bands (marked as L+2, L-2, R+2 and R-2) and a shadow band (marked as SB). {\textbf{b,}} Schematic Fermi surface of Bi2212. It consist of  the main Fermi surface (ML and MR), first-order superstructure Fermi surface (L+1, L-1, R+1 and R-1), second-order superstructure Fermi surface ( L+2, L-2, R+2 and R-2) and a shadow Fermi surface (SB).  {\textbf{c,}} Photoemission images obtained from \textbf{a} by converting the detector angle into the sample emission angle. Without the sample bias (U=0\,V), the sample emission angle equals to the detector angle (top panel). When the sample bias is applied (|U|=90\,V), the middle panel is obtained by using the position limit, while the lower panel is obtained by using the angular limit. {\textbf{d,}} Angle distribution curves (ADCs) at the Fermi level derived from \textbf{c}. The ADC peaks corresponding to the right and left second-order superstructure bands (R-2 and L-2) are marked. The position of the two peaks in the middle red curve obtained in the position limit match closely with the original two peaks in the top black curve (U=0\,V), while the position of the two peaks in the lower blue curve obtained in the angular limit shows a significant deviation. It validates the position limit in converting the detector angle into the emission angle. 
}
\label{fig7_positionlimit}
\end{figure*}

\begin{figure*}[tbp]
\begin{center}
\includegraphics[width=1.0\columnwidth,angle=0]{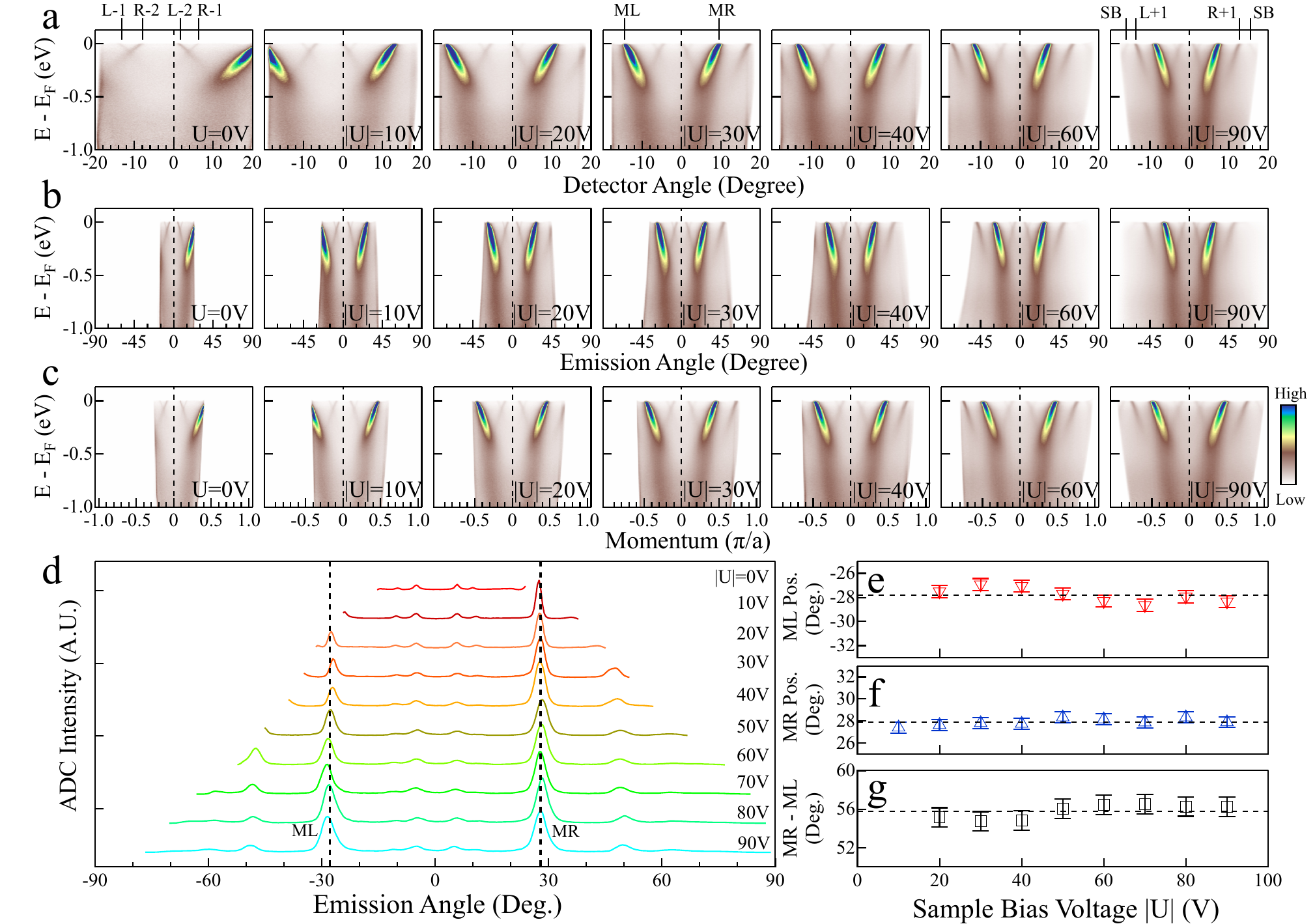}
\end{center}
\caption{\textbf{The increase of the covered emission angle and the corresponding momentum ranges with the application of bias on the sample.} {\textbf{a,}} Original photoemission images captured on the analyzer detector for Bi2212 at different sample bias voltages, with the position of the momentum cut indicated in fig.~\textbf{7b}. {\textbf{b,}} Photoemission images obtained from (a) plotted in the photoelectron emission angle. {\textbf{c,}} Photoemission images obtained from (b) plotted in the momentum space. It is clear that the covered emission angle and corresponding momentum range increase with the increasing sample bias voltage. {\textbf{d,}} Angle distribution curves (ADCs) at the Fermi level obtained from \textbf{b} measured at different bias voltages from $U=0\,V$ to $|U|=90\,V$. The ADC peaks corresponding to the left and right main bands are labelled as ML and MR, respectively. {\textbf{e-g,}} Extracted ADC peak positions of the ML peak (\textbf{e}), MR peak (\textbf{f}) and their position difference (\textbf{g}) obtained from \textbf{d}. 
}
\label{fig8_cut}
\end{figure*}

\begin{figure*}[tbp]
\begin{center}
\includegraphics[width=1.0\columnwidth,angle=0]{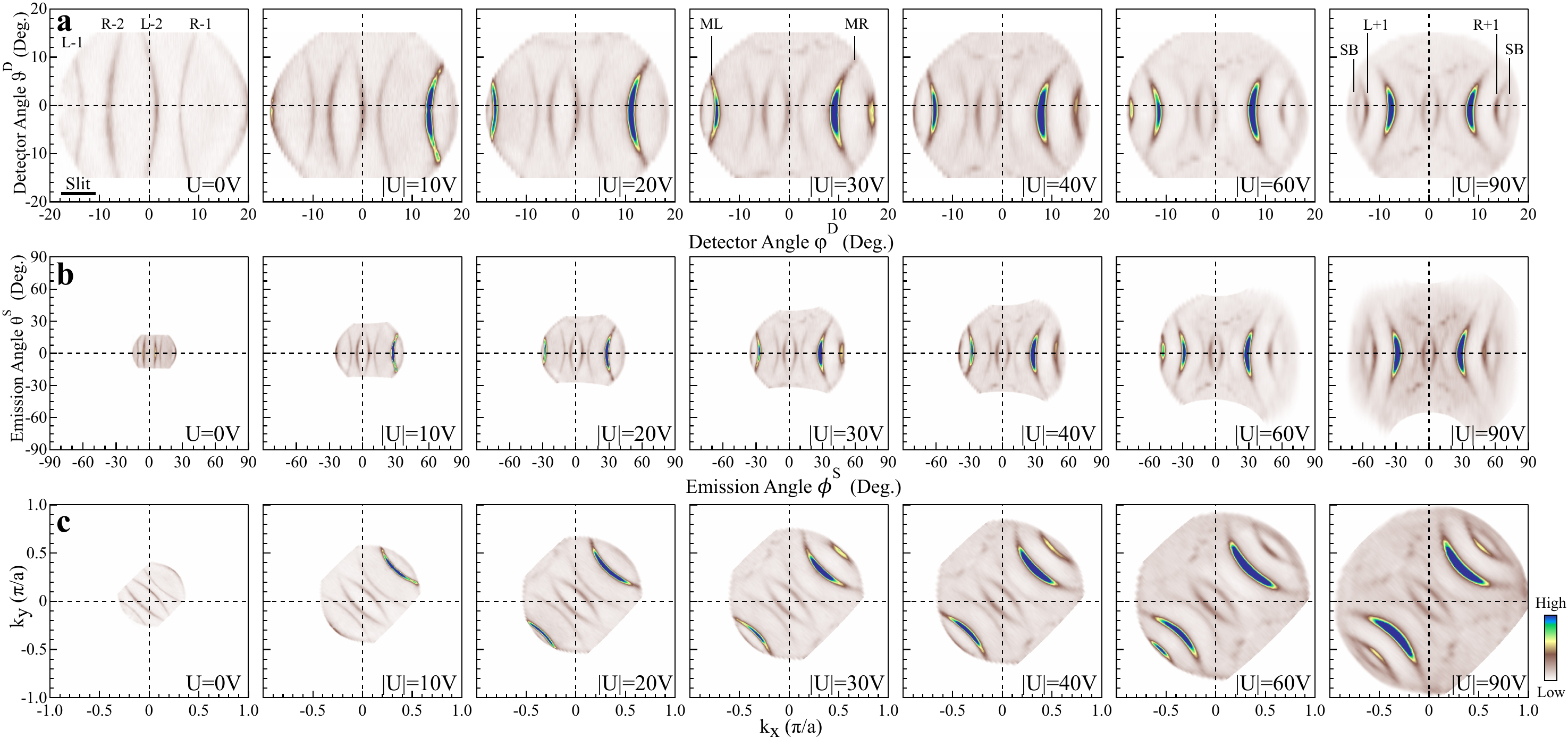}
\end{center}
\caption{\textbf{The increase of the covered two-dimensional emission angle range and the corresponding momentum space with the application of bias on the sample.}~{\textbf{a,}} Original distribution of the spectral weight at the Fermi level in the 2-dimensional analyzer detector angles measured on Bi2212 with different sample bias voltages. At zero bias, The 30-degree angular mode of our DA30L analyzer can cover [-19.6,19.6] angle range along the horizontal direction (defined as horizontal detector angle $\varphi^D$). It can also cover [-15,15] angle range along the vertical direction (defined as vertical detector angle $\vartheta^D$) without rotating the sample by using the DA30 mode to deflect photoelectrons along the vertical direction (leftmost panel). The measured images consist of the strong main Fermi surface sheets (ML and MR), weak first-order superstructure Fermi surface sheets (L+1, L-1, R-1 and R+1), second-order superstructure Fermi surface sheets (L+2, L-2, R-2 and R+2) and a shadow Fermi surface sheet (SB) as indicated in \cref{fig7_positionlimit}b. {\textbf{b,}} Distribution of the spectral weight obtained from (a) plotted in the photoelectron emission angles $\phi^S$ and $\theta^S$. {\textbf{c,}} Fermi surface mapping obtained from (b) plotted in the \{k$_x$,k$_y$\} momentum space. For comparison, the main Fermi surface (antibonding branch) is plotted in the first and third quadrants as black lines.
}
\label{fig9_FS}
\end{figure*}

\begin{figure*}[tbp]
    \begin{center}
    \includegraphics[width=1.0\columnwidth,angle=0]{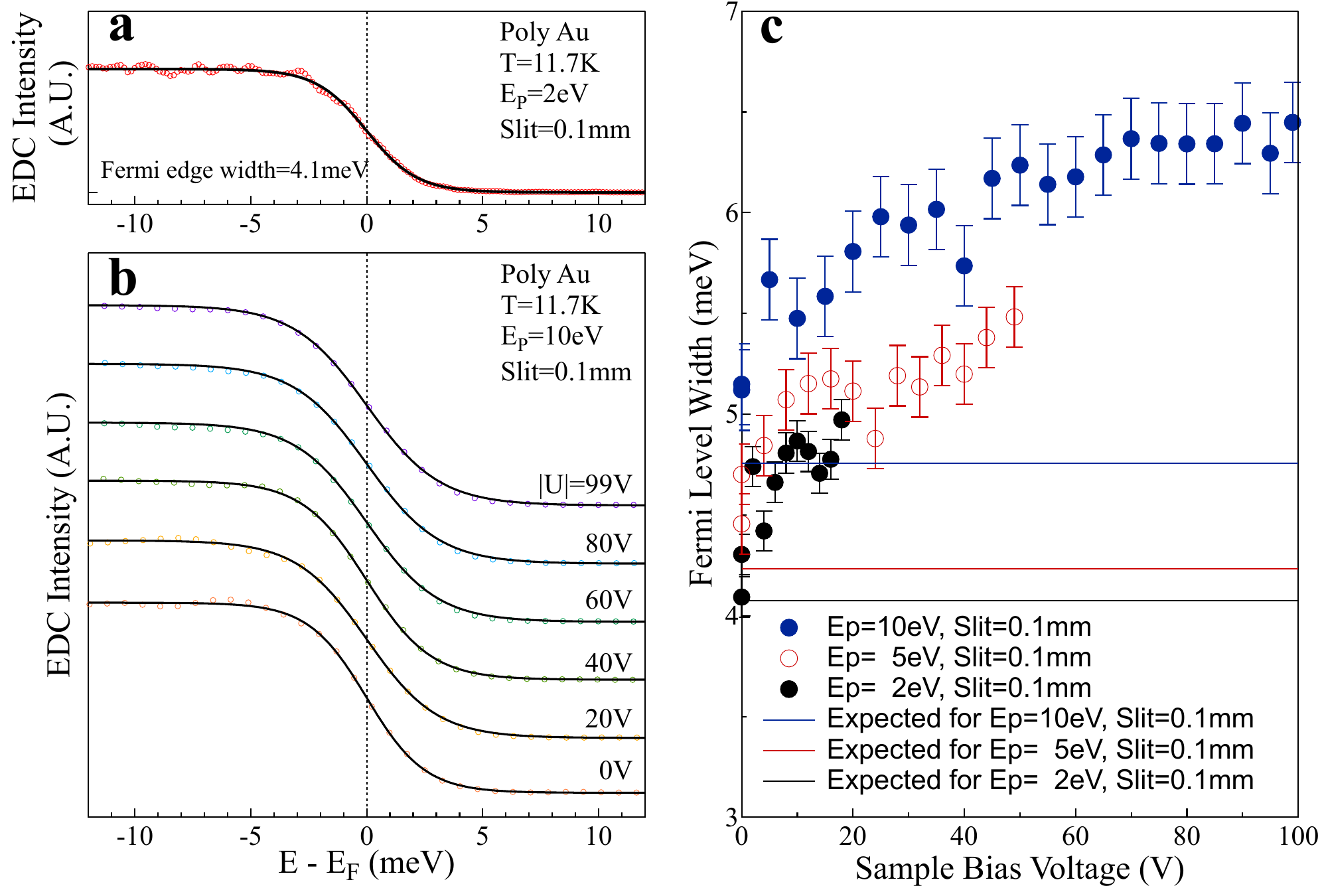}
    \end{center}
    \caption{\textbf{Effect of the Sample Bias on the Energy Resolution of Photoemitted Electrons.} {\textbf{a,}} Energy distribution curve (EDC) of polycrystalline gold measured  at 11.7\,K using 2\,eV pass energy and 0.1\,mm slit (open circles). No bias voltage was applied to the sample. The measured curve was fitted by the Fermi-Dirac distribution function (black solid line) and the fitted Fermi edge width (12-88\%) is 4.1\,meV. The corresponding energy resolution of the analyzer is 0.5\,meV and the linewidth of the laser is 0.26\,meV. {\textbf{b,}} EDCs of the polycrystalline gold measured at 11.7\,K at different bias voltages from $0V$ to $99V$, using 10\,eV pass energy and 0.1\,mm slit (open circles). The measured curves were fitted by the Fermi-Dirac distribution function (black solid lines) to get the Fermi edge width (12-88\%). {\textbf{c,}} The Fermi edge width measured at different bias voltages using different pass energies. The slit size is fixed at 0.1\,mm. The expected energy broadening, including the laser linewidth (0.26\,meV), thermal broadening of 11.7\,K (4.04\,meV) and analyzer with 10\,eV (2.5\,meV), 5\,eV (1.25\,meV) and 2\,eV (0.5\,meV), is plotted as blue, red and black solid lines, respectively. 
    }
    \label{fig10_energyresolution}
\end{figure*}

\begin{figure*}[tbp]
    \begin{center}
    \includegraphics[width=1.0\columnwidth,angle=0]{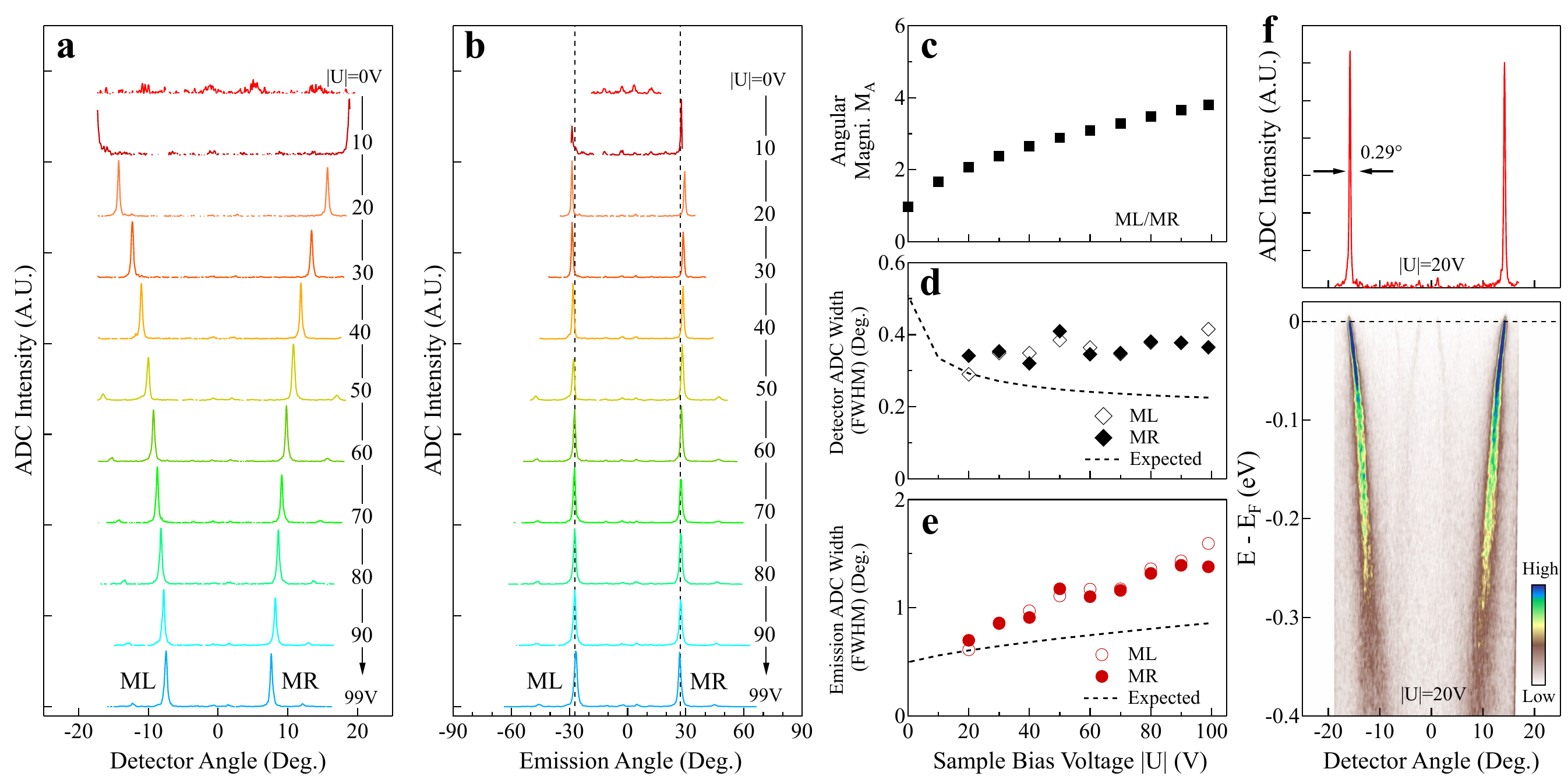}
    \end{center}
    \caption{\textbf{Effect of the Sample Bias on the Angular Resolution of Photoemitted Electrons.} {\textbf{a,}} Angular distribution curves (ADCs) at the Fermi level in the detector angle measured along the \(\Gamma-Y\) direction on the optimally-doped Bi2212 (\(T_c=91\,K\)) at 17\,K under different bias voltages from $0\,V$ to $99\,V$ using 30-degree angular mode. The two peaks corresponding to the ML and MR main bands are marked. {\textbf{b,}} ADCs in the emission angle obtained from \textbf{a} by converting the detector angle into the emission angle. {\textbf{c,}} Angular magnification factor (\(M_A\)) as a function of bias voltage for the ML and MR main bands obtained from \cref{fig4_magnification}. {\textbf{d,}} ADC width (full width at half maximum, FWHM) in the detector angle obtained from Lorentzian fitting the ML and MR peaks in \textbf{a}. The black dashed curve represents the expected values from \(\sqrt{(dA_S/M_A)^2+(dA_D)^2}\) by assuming the intrinsic ADC peak width of the sample \(dA_S=0.5^\circ\) and the angular resolution of the analyzer DA30 defection mode (\(dA_D=0.2^\circ\)).  {\textbf{e,}} The ADC width (FWHM) in the emission angle obtained from Lorentzian fitting the ML and MR peaks in \textbf{b}. The black dashed curve represents the expected values from \(\sqrt{(dA_S)^2+(M_A\times dA_D)^2}\) by assuming \(dA_S=0.5^\circ\) and (\(dA_D=0.2^\circ\)).  {\textbf{f,}} Band structure in the detector angle measured at a bias voltage of -20V (lower panel). The top panel shows the corresponding ADC at the Fermi level. A narrow peak with a width (FHWM) of \(0.29^\circ\) is observed.
    }
    \label{fig11_angularresolution}
\end{figure*}

\begin{figure*}[tbp]
    \begin{center}
    \includegraphics[width=1.0\columnwidth,angle=0]{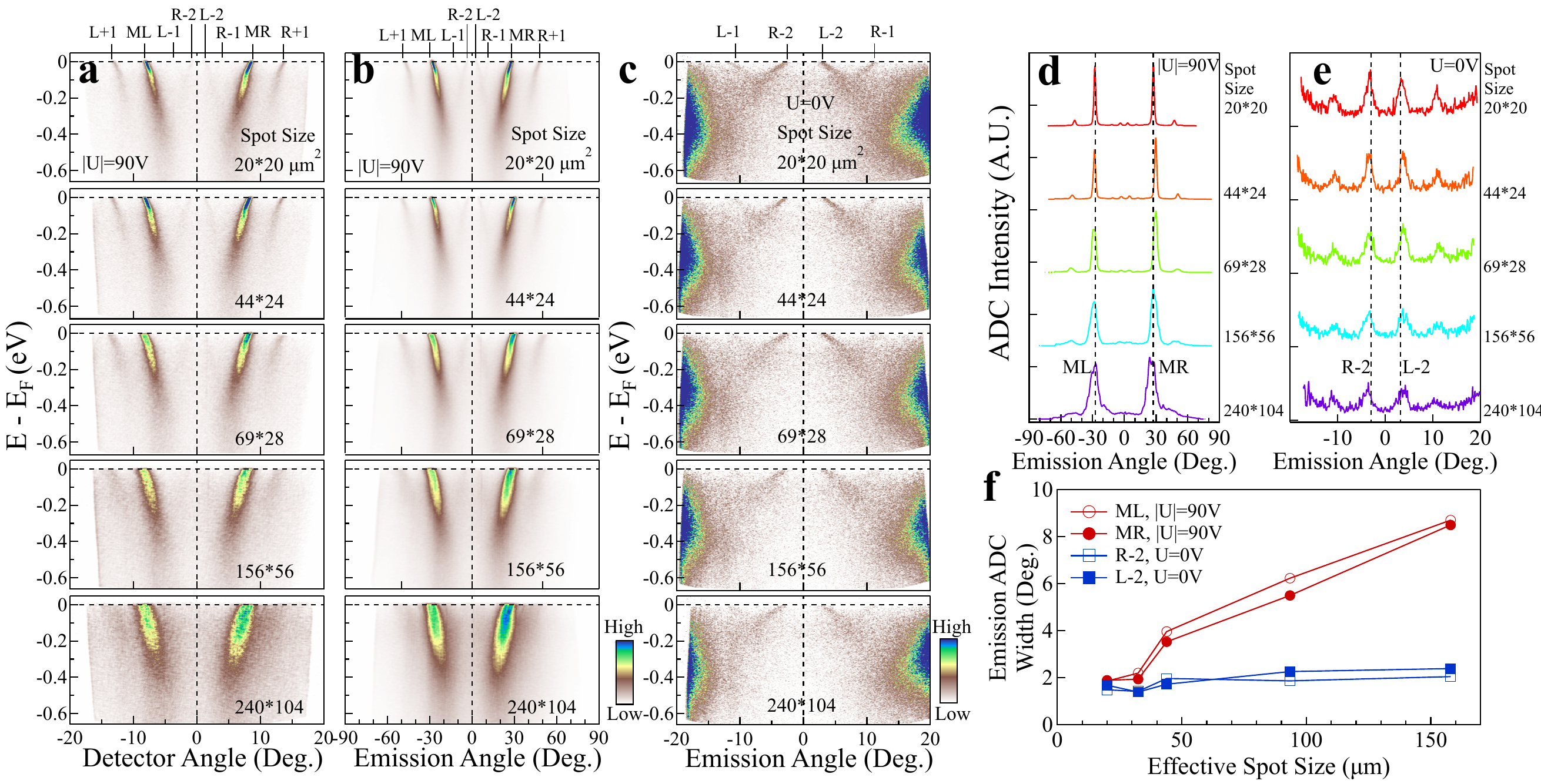}
    \end{center}
    \caption{\textbf{Effect of the Laser Beam Size on the Bias ARPES Measurements.}  {\textbf{a,}} Photoemission images on the analyzer detector measured on optimally-doped Bi2212 (\(T_c=91\,K\)) at 17\,K along \(\Gamma-Y\) under the bias voltage of 90\,V using different laser spot size from \(20\times 20\,\mu m^2\) to  \(240\times 104\,\mu m^2\). Here the size of the elliptical laser spot is defined as \(\sigma_x\times \sigma_z\) where \(\sigma_x\) represents the spot size along horizontal direction while \(\sigma_z\) represents the size along the vertical direction. The measured images consist of the strong main bands (marked as ML and MR), weak first-order superstructure bands (marked as L+1, L-1, R+1 and R-1) and second-order superstructure bands (marked as L+2, L-2, R+2 and R-2). {\textbf{b,}} Photoemission images in the emission angle obtained from \textbf{a} by converting the detector angle into the emission angle. {\textbf{c,}} Photoemission images in the emission angle measured under zero bias voltage but using different laser spot size. {\textbf{d,}} Angle distribution curves (ADCs) at the Fermi level obtained from \textbf{b} measured using different laser spot size. The ADC peaks corresponding to the left and right main bands are labelled as ML and MR, respectively. 
    {\textbf{e,}} ADCs at the Fermi level obtained from \textbf{c} measured using different laser spot size. The ADC peaks corresponding to the left and right second-order superstructure bands are labelled as R-2 and L-2, respectively. {\textbf{f,}} Extracted ADC peak width (FWHM) of the ML peak (open red circles) and MR peak (full red circles) from {\textbf{d}} and the R-2 peak (open blue squares) and L-2 peak (full blue squares) from {\textbf{e}} as a function of the effective spot size, defined as \(\sqrt{\sigma_x\times \sigma_z}\).
    }
    \label{fig12_beamsize}
\end{figure*}

\begin{figure*}[tbp]
\begin{center}
\includegraphics[width=0.8\columnwidth,angle=0]{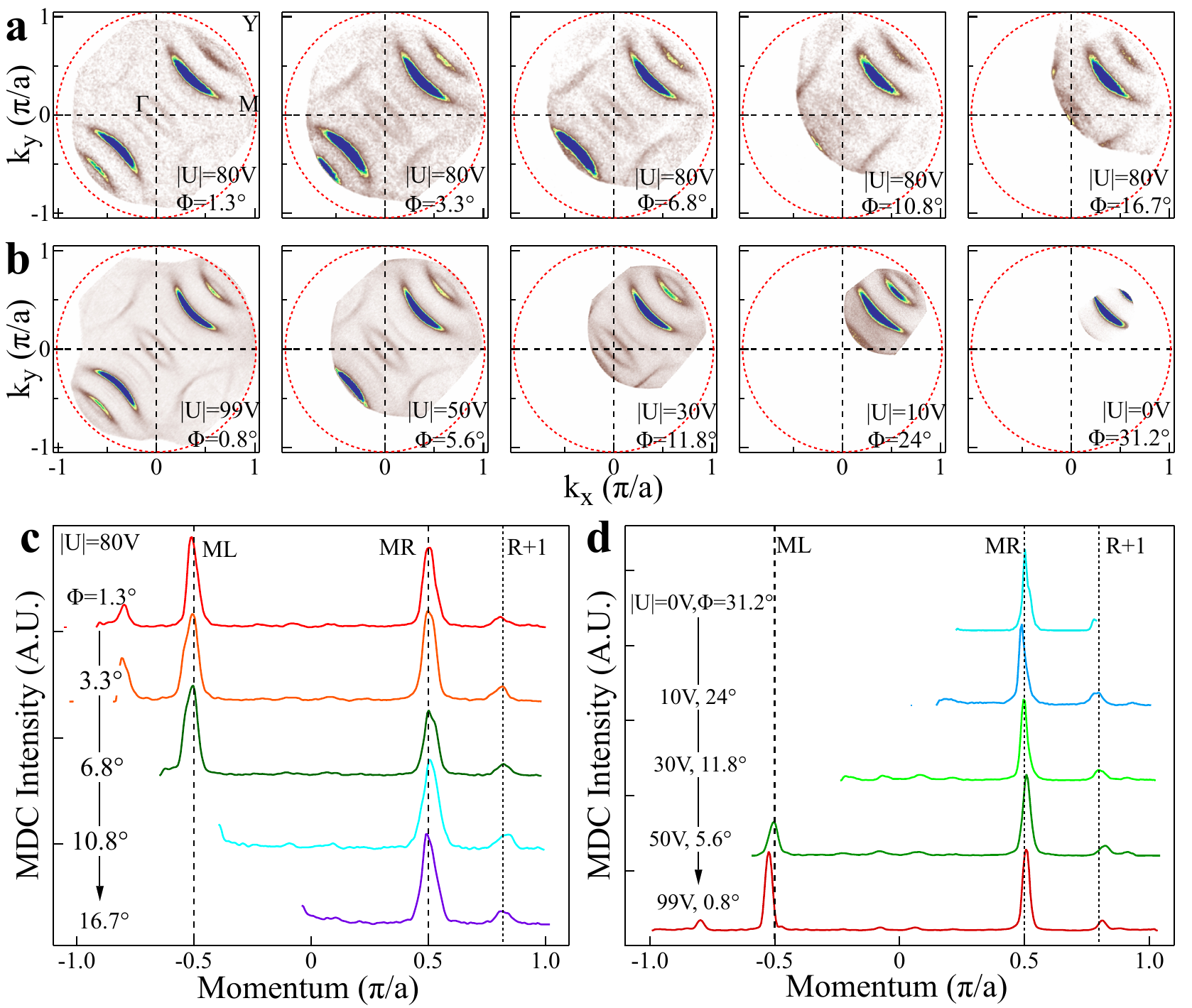}
\end{center}
\caption{\textbf{Performance Test of the Bias ARPES with the Sample being Oriented Off the Normal Emission.}~{\textbf{a,}} Fermi surface mappings of Bi2212 measured under a bias voltage of |U|=80\,V. The sample is tilted off the normal emission along \(\Gamma-Y\) by different angles from \(\Phi=1.3\) (leftmost panel) to \(\Phi=16.7^\circ\) (rightmost panel). The red dashed circle indicates the maximum momentum range corresponding to 2\(\pi\) solid angle  for the 6.994\,eV laser source. As the sample is tilted, the covered momentum space gradually shifts from four quadrants to the first quadrant. But the observed features remain similar in different measurements. {\textbf{b,}} Fermi surface mappings of Bi2212 measured under different combinations of the bias voltage and the sample tilt angle. With increasing tilt angle and the lowering of the bias voltage, the covered momentum space shifts from the initial four quadrants (leftmost panel) to the first quadrant (the third and fourth panels), but the right side of the covered momentum space remains close to the 2\(\pi\) solid angle limit even with a low bias voltage of 10\,V. The observed features remain similar in different measurements. {\textbf{c,}} Momentum distribution curves (MDCs) at the Fermi level, along the \(\Gamma-Y\) direction obtained from \textbf{a}. The two main bands (ML and MR) stay at similar positions in the measurements of different tilt angles. \textbf{d,} MDCs at the Fermi level, along the \(\Gamma-Y\) direction obtained from \textbf{b}. The two main bands (ML and MR) and the first-order superstructure band (R+1) stay at similar positions in the measurements of different tilt angles and sample bias voltages.
}
\label{fig13_rotation}
\end{figure*}

\begin{figure*}[tbp]
    \begin{center}
    \includegraphics[width=1.0\columnwidth,angle=0]{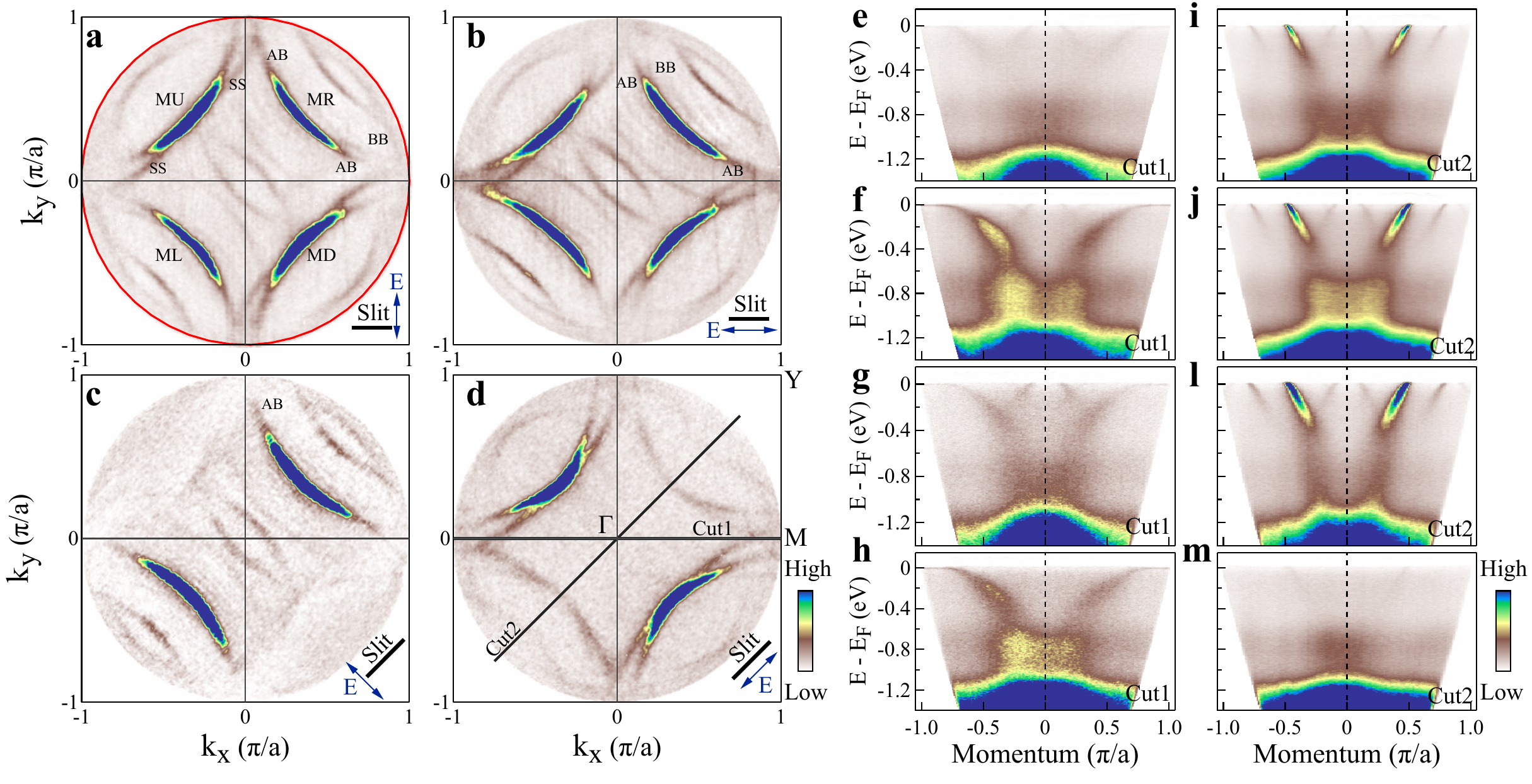}
    \end{center}
    \caption{\textbf{Full 2$\pi$ Solid Angle Collection of Photoelectrons in Measuring Bi2212 (Overdoped, T$_c$=67\,K) Using 6.994\,eV Laser by Applying a Sample Bias Voltage of 140\,V.} The work function of the sample is 4.377\,eV. {\textbf{a-d,}} Fermi surface mappings measured at 15\,K using 160\,V bias voltage under different polarization geometries by using the DA30 deflection mode. The direction of the electric field vector ($\vec{E}$) is indicated by blue double-headed arrow while the analyzer slit orientation is marked by black solid line in each panel. The entire 2$\pi$ solid angle of photoelectrons are collected and the maximum momentum can reach 1.002 $\pi/a$ which fully cover the $(\pi,0)$ and $(0,\pi)$ antinodal regions. The red solid circle represents the maximum momentum space that can be measured when the full $2\pi$ solid angle of photoelectrons is collected using  6.994\,eV laser. {\textbf{e-h,}} Band structures measured along the momentum Cut1 obtained from \textbf{a-d}, respectively. The location of the momentum Cut1 is indicated by a black line in \textbf{d}. {\textbf{i-m,}} Same as \textbf{e-h} but along the momentum Cut2. The location of the momentum Cut2 is indicated by another black line in \textbf{d}. 
    }
    \label{fig14_Bi2212}
\end{figure*}

\begin{figure*}[tbp]
    \begin{center}
    \includegraphics[width=1.0\columnwidth,angle=0]{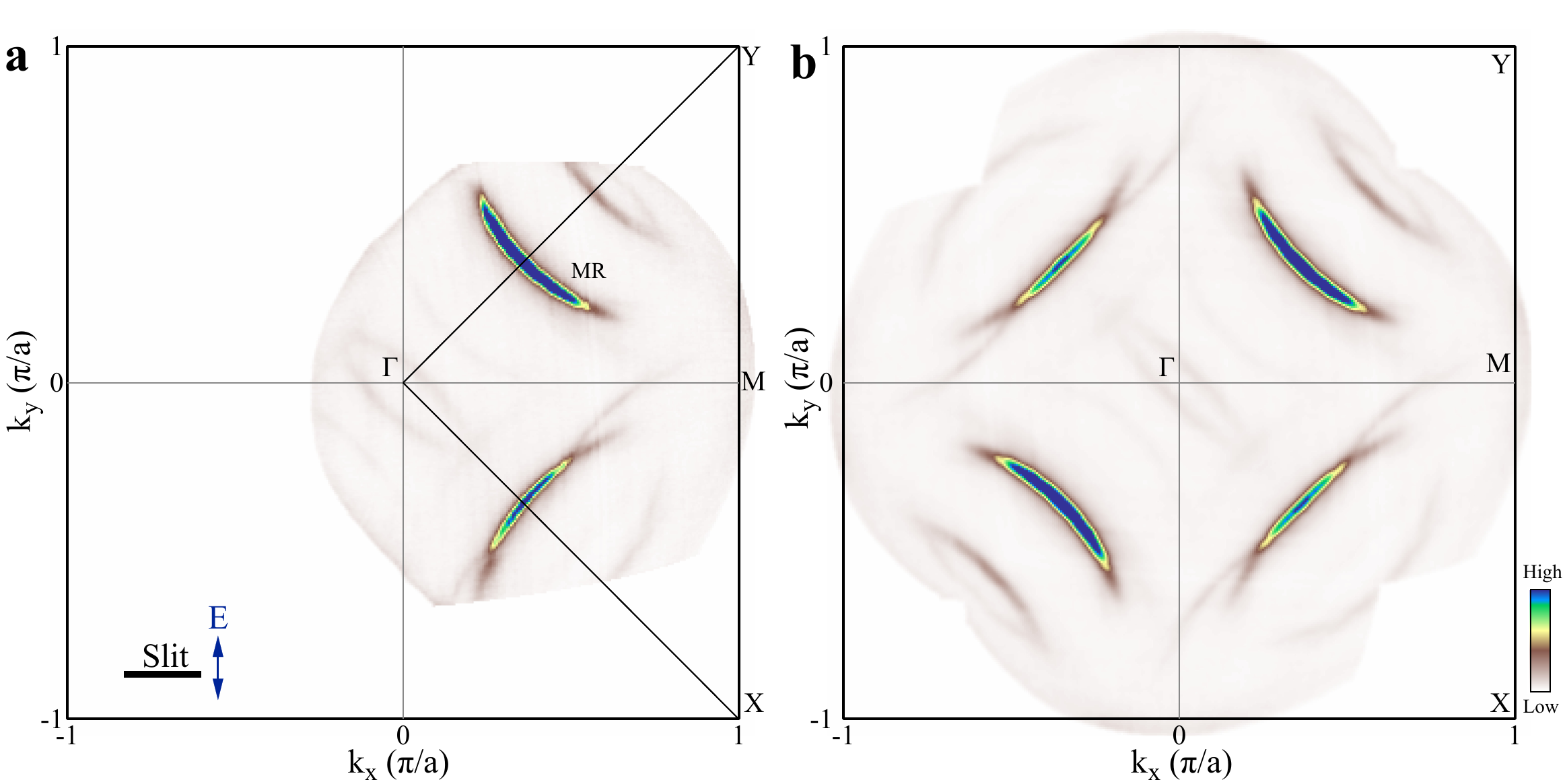}
    \end{center}
    \caption{\textbf{Fermi surface mapping of Bi2212 (optimally-doped, T$_c$=91\,K) Using a Sample Bias Voltage of 40\,V.} The work function of the sample is 4.154\,eV. {\textbf{a,}} Fermi surface mapping measured at 17\,K using 5\,eV pass energy and 0.1\,mm slit under a bias voltage of 40\,V by using the DA30 deflection mode. The analyzer slit is along \((0,0)\)-\((\pi,0)\) direction while the electric field vector ($\vec{E}$) is along \((0,0)\)-\((0 , \pi)\) direction. The sample is tilted along \((0,0)\)-\((\pi,0)\) direction by \(11.2\) degree. The covered momentum space can reach 1.04 $\pi/a$ on the right side. {\textbf{b,}} Symmetrized Fermi surface mapping obtained from \textbf{a} by taking the data enclosed by the \((0,0)\)-\((\pi,\pi)\) and \((0,0)\)-\((\pi,-\pi) \) lines and symmetrizing it using two mirror symmetries. 
    }
    \label{fig16_offnormalEp5}
\end{figure*}

\begin{figure*}[tbp]
    \begin{center}
    \includegraphics[width=1.0\columnwidth,angle=0]{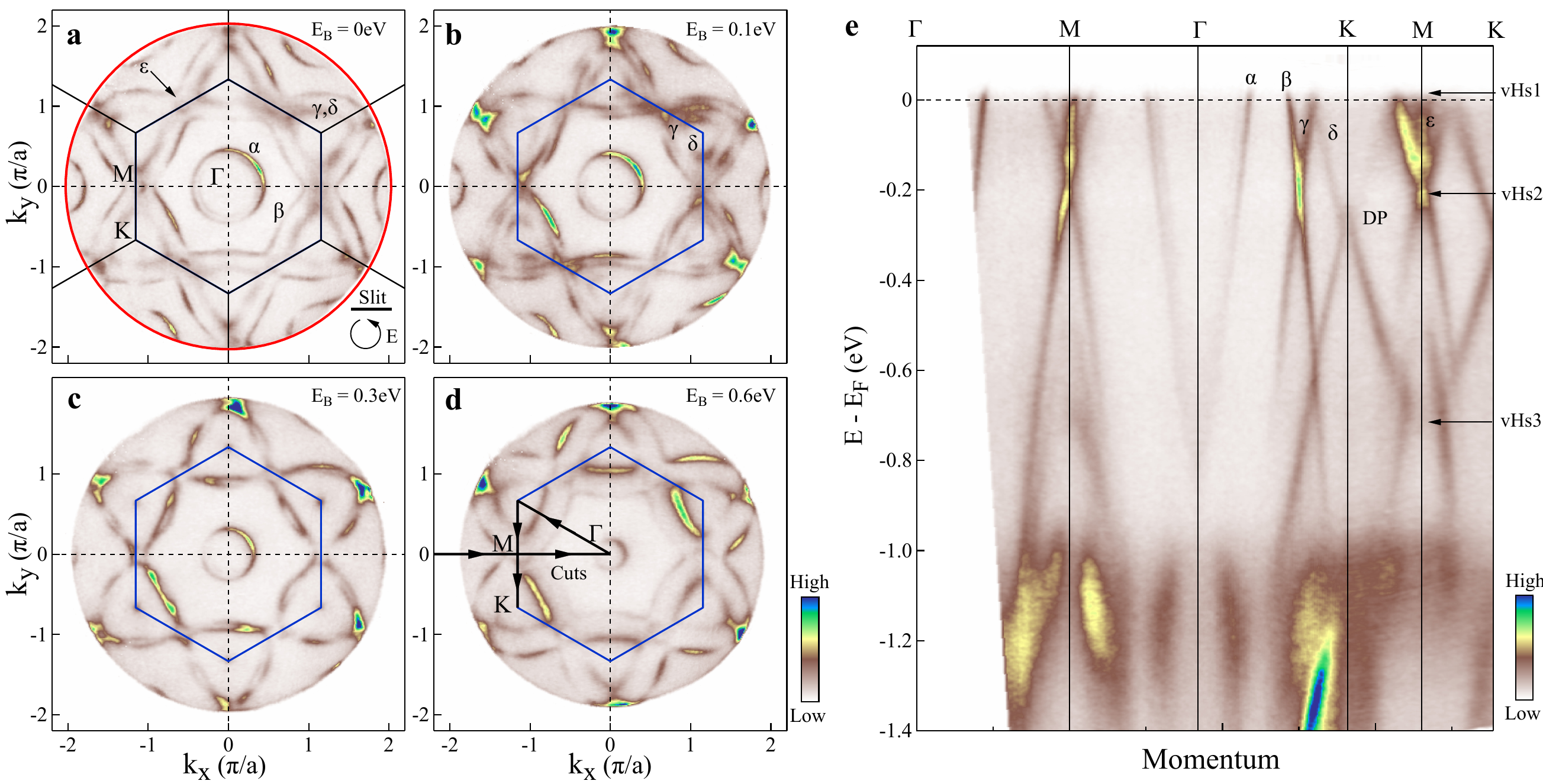}
    \end{center}
    \caption{\textbf{Full 2$\pi$ Solid Angle Collection of Photoelectrons in Measuring CsV\(_3\)Sb\(_5\) Using 6.994\,eV Laser by Applying a Sample Bias Voltage of 200\,V.} The work function of the sample is 1.345\,eV. {\textbf{a-d,}} Fermi surface (\textbf{a}) and constant energy contours at binding energies of 0.1\,eV (\textbf{b}), 0.3\,eV (\textbf{c}) and 0.6\,eV (\textbf{d}), measured at 94\,K using DA30 deflection mode with 20\,eV pass energy and 0.1\,mm slit and circularly polarized light. The Fermi surface mapping is obtained by scanning the vertical detector angle every 0.1$^\circ$ without shifting or rotating the sample and the acquisition time is about 1.5 hours. The measured momentum sapce covers the entire first Brillouin zone (blue hexagon) and reaches nearly the center of the second Brillouin zone. The red circle in \textbf{a} represents the maximum momentum space that can be measured when the full \(2\pi\) solid angle of photoelectrons is collected using 6.994\,eV laser. {\textbf{e,}} Band structures extracted along high-symmetry momentum cuts as marked by black lines in \textbf{d}
    }
    \label{fig15_CVS}
\end{figure*}

In actual APRES systems, it is inevitable that the electrical field environment around the sample may deviate from the ideal parallel plate capacitor model. This means that the bias experienced by the photoelectrons may be different from the applied voltage during the bias ARPES measurements and must be corrected. This effect appears to become more pronounced at high bias voltages when it approaches the critical value to collect the 2\(\pi\) solid angle of photoelectrons. We identified the correction term \(\eta\) to account for the deviation in the conversion relations. The value \(\eta\) can be determined from the photoelectron cone boundary (\cref{fig6_workfunctioncurve}). This effect is not intrinsic to the sample but depends on the particular sample and electrical field environments. Therefore, this correction must be determined under the exactly same experimental conditions that the bias ARPES measurements are carried out. The correction should also be made by considering the symmetry, the Brillouin zone boundary, the results with zero bias and the known electronic structures that have been determined before. These corrections are essential for accurately determining the electronic structures of the measured materials. 

There remains a plenty of room for improvements in the bias ARPES technique. In the case of our laser ARPES system, the energy resolution can be further optimized by minimizing the effect from the source meter. The angular resolution can be improved by using a smaller beam size and measuring the sample off normal emission to reduce the sample bias voltage. In particular, future efforts will be made to make the electric field around the sample more uniform and more consistent with the parallel-plate capacitor model. Moreover, the bias ARPES technique can be extensively used to expand the covered momentum space in ARPES measurements with different light sources including Helium discharge lamp, synchrotron radiation and other laser sources. In this respect, laser-based bias ARPES has its inherent advantage because the corresponding lower electron kinetic energy makes it easier to change the photoelectron trajectory to expand momentum space and reach the full 2\(\pi\) solid angle of photoelectrons at low bias voltage. The lower bias voltage is beneficial to preserve the high energy and angular resolutions.

In summary, the bias ARPES has added one more significant advantage to laser ARPES by dramatically expanding the momentum space and even reaching the full 2\(\pi\) solid angle collection of photoelectrons. In the meantime, it keeps all of the unique advantages of laser ARPES, including the ultrahigh energy resolution, fine momentum resolution, low sample temperature, enhanced bulk sensitivity and high data statistics. Such a technique is applicable to all the ARPES measurements with different light sources. The execution of bias ARPES is simple, robust, and easily integrated into existing ARPES platforms without substantial instrumental modifications, making it broadly applicable across a wide range of systems. This development makes bias ARPES, especially laser-based bias ARPES, more powerful and versatile in studying electronic structures of quantum materials.

\newpage

%{\bf Methods}
%{\bf ARPES.} 

\noindent {\bf Acknowledgement}\\
This work is supported by the National Key Research and Development Program of China (Grant Nos. 2021YFA1401800, 2022YFA1604200, 2022YFA1403900, 2023YFA1406002, 2024YFA1408301 and 2024YFA1408100), the National Natural Science Foundation of China (Grant Nos. 12488201, 12374066, 12374154 and 12494593),  the Innovation Program for Quantum Science and Technology (Grant No. 2021ZD0301800), CAS Superconducting Research Project (Grant No. SCZX-0101) and the Synergetic Extreme Condition User Facility (SECUF).\\

\noindent {\bf Author Contributions}\\
T.M.M. and X.J.Z. proposed and designed the research. T.M.M. carried out the ARPES experiments. T.M.M., Y.X., B.L., W.P.Z., N.C, M.K.X., D.W., H.Z.G., W.J.M., S.J.Z., F.F.Z., F.Y., Z.M.W., Q.J.P., Z.Y.X., Z.H.Z., X.T.L., M.H.Q., L.Z., G.D.L. and X.J.Z. contributed to the development and maintenance of Laser ARPES systems. T.M.M. and X.J.Z. analyzed the data and wrote the paper. All authors participated in discussions and comments on the paper.

% \bibliographystyle{naturemag_MTM}
% % % \bibliographystyle{apsrev4-1}
% \bibliography{references2} 

\newpage

\FloatBarrier

\begin{table}[ht]
  \centering
  \renewcommand{\arraystretch}{1.3} 
  \setlength{\tabcolsep}{5pt}       
  \caption{Work Functions of several Prototype Samples.}
  \begin{tabular}{|c|c|c||c|c|c|}
  \hline
  \textbf{Category} & \textbf{Samples} & \textbf{\shortstack{\\[0.3em]Work\\[0.2em]Function}} &\textbf{Category} & \textbf{Samples} & \textbf{\shortstack{\\[0.3em]Work\\[0.2em]Function}} \\
  \hline
  \multirow{6}{*}{\shortstack{\\[2.1em]Cuprate\\Superconductors}}   
                          & \shortstack{\\[0.3em]Bi2201\\[0.2em](OP, $T_c$=34K)} & 4.013\,eV &
  \multirow{8}{*}{\shortstack{\\[3.3em]Topological\\Materials}}
                         & Bi\(_2\)Se\(_3\) & 5.569\,eV\\
  \cline{2-3} \cline{5-6}
                          & \shortstack{\\[0.3em]Bi2212\\[0.2em](UD, $T_c$=80K)}& 3.896\,eV &
                          & Bi\(_2\)Te\(_3\) & 5.216\,eV \\
  \cline{2-3} \cline{5-6}
                         & \shortstack{\\[0.3em]Bi2212\\[0.2em](OP, $T_c$=91K)}  & 4.154\,eV &
                          & BiTeI & 5.318\,eV \\
 \cline{2-3} \cline{5-6}
                          & \shortstack{\\[0.3em]Bi2212\\[0.2em](OD, $T_c$=67K)}  & 4.377\,eV &
                          &  ZrTe\(_5\) & 4.660\,eV  \\
  \cline{2-3} \cline{5-6}
                          &  \shortstack{\\[0.3em]Bi2223\\[0.2em](OP, $T_c$=110K)}& 4.323\,eV &
                            & CsCr\(_6\)Sb\(_6\) & 2.505\,eV \\ 
  \cline{2-3} \cline{5-6}
                          &  \shortstack{\\[0.3em]Hg1223\\[0.2em](OP, $T_c$=134K)}& 3.711\,eV &
                            & \shortstack{\\Co\(_3\)Sn\(_2\)S\(_2\)\\(Sn termination)}  & 4.643\,eV\\
  \cline{1-3} \cline{5-6}
  \multirow{6}{*}{\shortstack{Iron-based\\Superconductors}}
                          & (Ba\(_{0.6}\)K\(_{0.4}\))Fe\(_2\)As\(_2\) & 1.947\,eV  &
                            & \shortstack{\\[0.3em]Co\(_3\)Sn\(_2\)S\(_2\)\\[0.2em](S termination)}& 5.267\,eV \\
  \cline{2-3} \cline{5-6}
                          & KFe\(_2\)As\(_2\) & 1.853\,eV  &
                          & MnBi\(_2\)Te\(_4\)  & 4.754\,eV  \\
  \cline{2-3} \cline{4-6}
                          &CsFe\(_2\)As\(_2\) & 1.173\,eV&
  \multirow{4}{*}{\shortstack{2-Dimensional \\ Materials}}
                          & 2H-NbSe\(_2\) & 6.051\,eV\\
  \cline{2-3} \cline{5-6}
                          & LiFeAs & 3.163\,eV &
                            & 1T'-WTe\(_2\)& 4.850\,eV\\
  \cline{2-3} \cline{5-6}
                          & FeSe/SrTiO\(_3\) & 4.816\,eV &
                          & 2H-MoS\(_2\)& 5.216\,eV\\
 \cline{2-3} \cline{5-6}
                          & FeSe\(_{0.4}\)Te\(_{0.6}\) & 4.848\,eV  &
  \multirow{3}{*}{\shortstack{\\[2.8em]Altermagnetism}}
                          & Co\(_{1/4}\)NbSe\(_2\) & 4.650\,eV \\
 \cline{1-3} \cline{4-6}
  \multirow{2}{*}{\shortstack{Nickelate\\Superconductors}}
                          & La\(_3\)Ni\(_2\)O\(_7\) & 3.860\,eV &
                          &KV\(_2\)Se\(_2\)O& 2.020\,eV \\
 \cline{2-3} \cline{5-6}
                          & La\(_4\)Ni\(_3\)O\(_{10}\) & 3.643\,eV  &
                          & Fe\(_3\)GeTe\(_2\) & 4.708\,eV\\
 \cline{1-3} \cline{4-6}
  \multirow{3}{*}{\shortstack{\\[2.8em]Other\\Superconductors}}
                          & \shortstack{\\[0.3em]CsV\(_3\)Sb\(_5\)\\[0.2em](Cs termination)} & 1.345\,eV &
  \multirow{4}{*}{\shortstack{Other\\Materials}} 
                        & \shortstack{\\[0.3em]Polycrystalline\\[0.2em]Au} & 5.286\,eV\\
  \cline{2-3} \cline{5-6}
                          & Sr\(_2\)RuO\(_4\) & 2.701\,eV &
                          & \shortstack{\\[0.3em]Polycrystalline\\[0.2em]Ag} & 4.351\,eV \\
  \cline{2-3} \cline{5-6}
                        &CeCoIn\(_5\) & 3.890\,eV &
                        & CeSb\(_2\)  & 4.433\,eV \\
  \cline{2-3} \cline{5-6}
                          &UTe\(_2\) & 4.498\,eV&
                          & PtBi\(_2\)& 4.211\,eV\\
  \hline 
  \end{tabular}
  \label{table_workfunction}
\end{table}

\end{document}